\documentclass[11pt,a4paper]{article}
\usepackage{jcappub}

 \usepackage{amsmath}
 \usepackage{float}
 \usepackage{placeins}
\usepackage{epsfig}
 
 \usepackage{graphicx}
 \usepackage{amssymb}
 \usepackage{amsfonts}
 \usepackage{latexsym}
 \usepackage[latin1]{inputenc} 

\renewcommand{\vec}[1]{{\bf #1}}

\def\beq{\begin{eqnarray}}
\def\eeq{\end{eqnarray}}
\def\bea{\begin{eqnarray}}       
\def\eea{\end{eqnarray}}
\def\ln{\,\mbox{ln}\,}

\def\al{\alpha}
\def\be{\beta}

\def\de{\delta}

\def\na{\nabla}

\def\si{\sigma}
\def\om{\omega}
\def\ph{\varphi}

\def\Ga{\Gamma}
\def\De{\Delta}
\def\La{\Lambda}

\title{Gravitational waves and stability of cosmological 
solutions in the theory with anomaly-induced corrections}

\author[a]{J\'{u}lio C. Fabris}
\author[b]{Ana M. Pelinson} 
\author[c]{Filipe de O. Salles}
\author[c,1]{and Ilya L. Shapiro \note{On leave from Tomsk State
Pedagogical University, Tomsk, Russia.}}

\affiliation[a]{Departamento de F\'{\i}sica, CCE,
Universidade Federal do Esp\'{\i}rito Santo, ES, Brazil}
\affiliation[b]{Departamento de F\'{\i}sica, CFM,
Universidade Federal de Santa Catarina, SC, Brazil}
\affiliation[c]{Departamento de F\'{\i}sica, ICE, 
Universidade Federal de Juiz de Fora, MG, Brazil}

\emailAdd{fabrisjc@yahoo.com.br}
\emailAdd{ana.pelinson@gmail.com}
\emailAdd{fsalles@fisica.ufjf.br}
\emailAdd{shapiro@fisica.ufjf.br}

\abstract{The dynamics of metric perturbations is explored in the gravity 
theory with anomaly-induced quantum corrections. Our first purpose 
is to derive the equation for gravitational waves in this theory on 
the general homogeneous and isotropic background, and then verify 
the stability of such background with respect to metric perturbations.
The problem under consideration has several interesting applications. 
Our first purpose is to explore the stability of the classical 
cosmological solutions in the theory with quantum effects taken 
into account. There is an interesting literature about stability 
of Minkowski and de Sitter spaces and here we extend the consideration
also to the radiation and matter dominated cosmologies. Furthermore, 
we analyze the behavior of metric perturbations during inflationary 
period, in the stable phase of the Modified Starobinsky inflation.}


\arxivnumber{number}

\begin{document}
\maketitle

\section{Introduction}

The semiclassical approach to gravity is usually associated 
with the equation 
\beq
G_{\mu\nu}\,=\,R_{\mu\nu} -\frac12\,Rg_{\mu\nu} \,=\,<T_{\mu\nu}>
\label{semi}
\eeq
and implies that the gravity itself is not quantized. The 
averaging in the {\it r.h.s.} of the last equation comes 
from the quantum matter fields. Different from what one can 
think, the {\it r.h.s.} may be very nontrivial even if no 
matter sources are present in the given point of space-time.
The reason is that the average of the Energy-Momentum tensor of the 
{\it vacuum} can be nontrivial, for it may depend on the curvature 
tensor components and its derivatives, with possible non-local 
structures. There are two subtle points in the equation 
(\ref{semi}). Let us start from the terminology. The 
renormalizable theory of matter fields on curved space-time 
background requires that the action of gravity should be 
extended compared to the one of General Relativity (GR)
\cite{UtiDW-62,birdav,book} (see also \cite{PoImpo} for 
the recent review). 
The full action includes Einstein-Hilbert term, which is 
the origin of the {\it r.h.s.} of (\ref{semi}) with the 
cosmological constant term 
\beq
S_{EH}
\,=\,-\,\frac{1}{16\pi G}\int d^4 x\sqrt{-g}\,\left(\,R+2\La\,\right)
\label{EH}
\eeq
and also the higher derivative terms 
\beq
S_{HD} &=& \int d^4x \sqrt{-g}
\left\{a_1C^2+a_2E+a_3{\Box}R+a_4R^2 \right\},
\label{HD}
\eeq
where $\,C^2=R_{\mu\nu\al\be}^2 - 2 R_{\al\be}^2 + (1/3)\,R^2\,$
is the square of the Weyl tensor and 
$\,E = R_{\mu\nu\al\be}^2 - 4 R_{\al\be}^2 + R^2\,$
is the integrand of the Gauss-Bonnet topological term. 
All terms of the action of vacuum 
\beq
S_{vac} &=& S_{EH}\,+\,S_{HD}\,,
\label{vacuum}
\eeq
belong to the gravitational action. If we consider the Einstein 
equations as something intended to define the relation between 
geometry and distribution of matter, it is clear that the whole 
$S_{vac}$ should contribute to the {\it l.h.s.} of (\ref{semi}). 
However, by traditional virtue we use to put all the contributions 
beyond the Einstein tensor to the {\it r.h.s.}. To a great extent 
this way of settling the terms is explained by the fact that Eq. 
(\ref{semi}) with the \ {\it classical} \ Energy-Momentum tensor 
at the {\it r.h.s.} work pretty well and provide very good fit 
for many observational tests of GR. So, it 
looks like all extra terms, except the cosmological 
constant, are in fact unnecessary at the classical level 
and therefore their introduction is no justified in a 
purely gravitational framework. In other words, why should 
the gravitational physicist who works with the very large 
scale phenomena, worry about the quantum notions, such 
as renormalizability? In fact, the use of the terms (\ref{HD})
may lead to serious problems, because these terms are known 
to produce unphysical ghost terms for the linearized 
gravitational field on the flat background \cite{Stelle-1978}. 
So, it is somehow unclear how to apply the consistent at 
quantum level theory (\ref{vacuum}) for the classical 
gravitational purposes. 

Another aspect of the same problem is related to quantum 
corrections to (\ref{vacuum}). In general, the problem of 
deriving such corrections is unsolved, but there is one 
important case where the situation is quite clear. The 
one-loop effective action of massless conformal fields is 
essentially controlled by conformal anomaly \cite{duff77}. 
Indeed, the anomaly-induced effective action of vacuum 
includes an arbitrary conformal functional, but in many 
particular cases its role is known to be very restricted. 
For instance, this functional is a trivial constant for the 
homogeneous and isotropic metric, where the anomaly-induced
quantum corrections produce Starobinsky inflation \cite{star}.
In the case of black holes, taking into account the conformal 
anomaly enables one to calculate Hawking radiation 
\cite{ChrFull} and using the anomaly-induced action 
one can even classify the vacuum states in the vicinity 
of the black hole \cite{balsan,And-Mot-RN}. The equations 
for gravitational waves calculated by using direct methods 
\cite{star83,HHR} and  anomaly-induced action \cite{wave} 
produce equivalent results at least on the de Sitter 
background. In both last cases the mentioned conformal 
functional apparently plays no role, which can be explained 
by the fact that the conformal anomaly picks up all quantum 
effects which correspond to the UV limit and, therefore, 
the remaining terms can be relevant only for the sub-leading 
effects. So, from the Quantum Field Theory viewpoint the 
anomaly-induced effective action of vacuum represents a 
well-defined quantum contribution which can be used to 
verify the compatibility of gravity and quantum effects in 
the sense we have discussed above. The late Universe 
represents, in fact, a very good opportunity for such a 
verification. The typical energy scale of gravity is given 
by the Hubble parameter or by the scale of the gravitational 
waves, both of which are much smaller than the masses of all 
quantum massive fields. Certainly, all such 
fields strongly decouple in the higher derivative sector 
of the theory \cite{apco} and therefore the only active 
quantum field is photon. So, we arrive at the conclusion 
that the anomaly-induced action of vacuum coming from the 
photon field is a safe approximation for the quantum 
contribution in the late Universe. The purpose of the 
present work is to verify whether these quantum terms are 
compatible with the well-known classical cosmological 
solutions for the different epochs of the history of the 
Universe. The method which will be used in what follows 
is based on the derivation of metric perturbations over 
the given cosmological solution, in the theory with 
anomaly-induced effective action of gravity. 

The investigation of stability of the physically important 
solutions in a more general gravitational theories has a long 
and interesting history, starting from \cite{Ranjbar}, where 
the stability of Minkowski quantum vacuum has been explored 
for the first time (see also \cite{BroutGunzig}). Similar program 
for the theory of quantum gravity at non-zero temperature has 
been carried out in \cite{Gross-82}. The stability of Minkowski 
has been further studied in \cite{AndMot-stab-2003}. The 
stability of de Sitter space for both semiclassical theory and 
quantum gravity, within different quantization schemes has been 
recently considered in \cite{Verdaguer}. Furthermore,
the stability of de Sitter space has been recently discussed 
in \cite{Burgess} by an unconventional method of power counting 
in the IR limit, with an apparently negative result concerning 
the validity of the whole semiclassical approximation (see 
further references therein). 
Compared to the methods of exploring stability used in the 
papers mentioned above, the anomaly-induced effective action 
has two important advantages: safety and simplicity. As we 
have already mentioned above, the anomaly-induced action is
picks up the most important non-local 
part of effective action. As far as it is based on the conformal 
anomaly of massless fields, it is closely related to both UV 
and IR limits of the theory and hence is independent on whether 
we use usual in-out or more complicated in-in formalism. Also 
(see next section) the anomaly-induced action is very simple, 
it is given by a compact and explicit non-local expression which 
can be easily made local by introducing two auxiliary scalar 
fields. Historically, the anomaly-induced effective action was 
the theoretical basis of the first cosmological models with 
quantum corrections \cite{fhh}, Starobinsky inflation \cite{star} 
and first derivation of cosmic perturbations in inflationary model 
\cite{much}. As we will see below, the use of anomaly-induced 
action enables one to explore stability not only for Minkowski or 
de Sitter spaces, but for a wider set of classical cosmological 
solutions. 

In our investigation of stability we will restrict consideration 
by the gauge invariant part of metric perturbations related to 
the gravitational wave. The solutions of our interest include 
the radiation and matter - dominated Universes, the late Universe 
dominated by the cosmological constant and the stable phase of the 
modified Starobinsky inflation \cite{Shocom,asta}. 

The paper is organized as follows.  In Sect. 2 present a 
brief review of anomaly-induced action of gravity and of 
the corresponding cosmological solutions. Sect. 3 is devoted 
to the derivation of gravitational waves equation on an 
arbitrary cosmological (homogeneous and isotropic) background. 
In Sect. 4 we explore the stability of classical solutions 
by means of approximate analytic method and in Sect. 5 
discuss the spectrum of metric perturbations. Sect. 6 is 
about the behavior of gravitational waves in the Modified
Starobinsky model of inflation. Finally, in the last 
section we draw our conclusions.

\section{Effective action induced by anomaly}

The covariant form of anomaly-induced effective action of 
gravity \cite{rie,frts84} is the most complete available form 
of the quantum corrections to the gravitational action in four 
space-time dimensions. The application to cosmology has been 
considered in \cite{fhh,Mamaev:1980nj} and led to the well-known 
Starobinsky model of inflation \cite{star} (see also more 
detailed description in \cite{vile} and consequent development 
in form of Modified Starobinsky model \cite{Shocom,asta}). 

The anomalous trace of the energy momentum tensor is given by 
the expression \cite{duff77,birdav}
\beq
<T_\mu^\mu> \,=\, - \,(wC^2 + bE + c{\Box} R)\,,
\label{trace an}
\eeq
where the coefficients $w,b$ and $c$ depend on the number of
active quantum fields of different spins,
\beq
w &=& \frac{1}{(4\pi)^2}\,\Bigg(
\frac{N_0}{120} + \frac{N_{1/2}}{20} + \frac{N_1}{10} \Bigg)\,,
\label{w}
\\
b &=&-\, \frac{1}{(4\pi)^2}\,\Bigg( \frac{N_0}{360} 
+ \frac{11\,N_{1/2}}{360} + \frac{31\,N_1}{180}\Bigg)\,,
\label{b}
\\
c &=& \frac{1}{(4\pi)^2}\,\Bigg( \frac{N_0}{180} + \frac{N_{1/2}}{30}
- \frac{N_1}{10}\Bigg) \,.
\label{c}
\eeq

It is easy to see that these coefficients are nothing else but the 
$\be$-functions for the parameters $a_{1,2,3}$ in the classical 
action of vacuum (\ref{vacuum}). Due to the decoupling phenomenon 
\cite{apco}, the number of active fields can vary from one 
epoch in the history of the Universe to another. As we have 
already mentioned in the Introduction, the present-day Universe 
corresponds to the particle content with $\,N_0 = N_{1/2}= 0\,$ 
and $\,N_1=1$.

The anomaly-induced effective action ${\bar \Ga}_{ind}$ 
represents an addition to the classical action of gravity, 
and can be found by solving the equation 
\beq
\frac{2}{\sqrt{-g}}\,g_{\mu\nu}
\frac{\de\, {\bar \Ga}_{ind}}{\de g_{\mu\nu}}
\,=\, <T_\mu^\mu> \,=\,(\om C^2 + bE + c{\Box} R)\,.
\label{mainequation}
\eeq
The covariant and generally non-local solution can be easily 
found in the form 
\beq
{\bar \Ga} &=& S_c[g_{\mu\nu}]\,-\,\frac{3c+2b}{36}\,
\int d^4 x \sqrt{-g (x)}\,R^2(x) 
\label{nonloc}
\\
&^+& \int d^4 x \sqrt{-g (x)}\, \int d^4 y \sqrt{-g (y)}\,
(E - \frac23{\Box}R)_x \,G(x,y)\,\left[\,\frac{w}{4}\,C^2
- \frac{b}{8}\,(E - \frac23{\Box}R)\right]_y \,,
\nonumber
\eeq
where $\,G(x,y)\,$ is a Green function for the operator
$$
\De_4 = {\Box}^2 + 2\,R^{\mu\nu}\na_\mu\na_\nu - \frac23\,R{\Box}
+ \frac13\,(\na^\mu R)\na_\mu\,.
$$
Finally, one can rewrite (\ref{nonloc}) in the local form 
by introducing two auxiliary fields $\phi$ and $\psi$
\cite{a} (see also \cite{Mottola-98} for an alternative 
albeit equivalent scheme),  
\beq
{\bar \Gamma}_{ind} 
&=& 
S_c[g_{\mu\nu}] - \frac{3c+2b}{36}\,
\int d^4 x \sqrt{-g (x)}\,R^2(x)
+  \int d^4 x \sqrt{-g (x)}\,\Big\{
 \frac12\ph\De_4\ph - \frac12\psi\De_4\psi 
\nonumber
\\
&+& \ph\,\Big[\,\frac{\sqrt{-b}}{2}\,\big(E -\frac23\,{\Box}R\big)\,
- \frac{w}{2\sqrt{-b}}\,C^2\,\Big]
+ \frac{w}{2 \sqrt{-b}}\,\psi\,C^2 \Big\}\,.
\label{finaction}
\eeq
The expression (\ref{finaction}) 
is classically equivalent to (\ref{nonloc}), because
if one uses the equations for the auxiliary fields $\,\ph\,$ and
$\,\psi$, the nonlocal action (\ref{nonloc}) is restored. 
Consider now the background cosmological solution for the theory 
with the action including quantum corrections,
\beq
S_{total}\, =\, -\, M^2_P\,\int d^4x\,\sqrt{-g}\,R\,
+ \,{\bar \Ga}\,,
\label{tota}
\eeq
where $M^2_P = {1}/{16\pi G}$ is the square of the
Planck mass, and
the quantum correction $\,{\bar \Ga}\,$ is taken in
the form (\ref{finaction}).

Before we start to deal with the application of the higher 
derivative theory (\ref{vacuum}) with the quantum correction
(\ref{nonloc}), let us make come general observations. First
of all, (\ref{vacuum}) and in particular (\ref{HD}) represent 
the classical action of external gravitational field. One has 
to introduce the higher derivative terms into the classical 
action, for otherwise the same terms emerge as divergences 
and can not be controlled by renormalization (see, e.g., 
\cite{birdav,book}).

One of the possibilities of how to deal with the higher 
derivative terms is to treat them as perturbations. This idea 
comes from QED, where the higher derivative terms are always 
small loop corrections and another approach leads to the 
run-away solutions. However, in the present case this is 
not necessary a right approach, exactly because (\ref{HD}) 
is part of the classical action and therefore should be 
treated at the same level as Einstein-Hilbert term and the
cosmological constant. In this case there will be nothing 
wrong with all known classical solutions of general relativity. 
The reason is that, due to the Planck suppression, the effect 
of these terms on Newton law, the Schwarzchild solution or on 
the evolution of the conformal factor in cosmology would be 
negligible. In fact, the effect of (\ref{HD}) in all these 
cases is exactly zero, which is obvious for cosmology and was 
recently explained in \cite{HDBH} for the black hole case. 
The situation is very much different for perturbations around 
these solutions, especially for the tensor mode, because the
last does not depend on the gauge-fixing. Now, the parameters 
of the action (\ref{HD}) are arbitrary and can be fixed in such 
a way that the given classical solution which is physically 
relevant would be stable. Different from the classical part, 
the non-local quantum term (\ref{nonloc}) is fixed almost 
completely (up to the conformal term). Therefore it is 
interesting to check whether the conditions of stability 
for the parameters of the classical action (\ref{HD}) do 
change or not in the presence of this well-defined quantum 
term. This task is exactly what we will try to accomplish 
in the rest of the paper. 

Let us start from the background cosmological metric. 
Looking for the isotropic and homogeneous solution, the 
starting point is to choose the metric in the
form $g_{\mu\nu} = a^2(\eta)\,{\bar g}_{\mu\nu}$, where
$\,\eta\,$ is conformal time. It proves useful to introduce the 
notation $\,\si = \ln a$. The theory includes the equations 
for the three fields, namely for $\,\ph,\,\psi,\,$ and $\,\si$. 
For the sake of simplicity we will consider conformally flat 
background and therefore set $\,{\bar g}_{\mu\nu} = \eta_{\mu\nu}$.

Equations for $\,\ph\,$ and $\,\psi\,$ have especially
simple form
\beq
&& \sqrt{-g}\,\left[\,\De_4\,\ph
+ \frac{\sqrt{-b}}{2}\,(E -\frac23\,{\Box}R)\,
- \frac{w}{2\sqrt{-b}}\,C^2\,\right] = 0\,,
\nonumber
\\
&& \sqrt{-g}\,\left[\,
\De_4\,\psi - \frac{w}{2\sqrt{-b}}\,C^2\,\right] = 0\,.
\nonumber
\eeq
By using the transformation laws for the quantities in the 
last expression, one can obtain
\beq
\sqrt{-g}C^2 = \sqrt{-{\bar g}}{\bar C}^2
\,,\,\,\,\,\,\,\,\,\,\,\,\,\,\,\,\,\,\,
\sqrt{- {g}}\,{\De}_4 = \sqrt{-{\bar g}}\,{\bar \De}_4 \,\,,
\label{weyly}
\eeq
\beq
 \sqrt{-g}(E - \frac23{\Box}R) =
\sqrt{-{\bar g}}({\bar E} - \frac23{\bar {\Box}}{\bar R}
+ 4{\bar {\De}}_4\si )\,.
\label{trans}
\eeq
Taking into account our choice for the fiducial metric
 $\,{\bar g}_{\mu\nu} = \eta_{\mu\nu}$ all the terms in the 
{\it r.h.s.} of the last equation are equal to zero and we 
arrive at the following equations
\beq
{\Box}^2\,\ph
+ 8\pi\sqrt{-b}\,{\Box}^2\si = 0\,,
\,\,\,\,\,\,\,\,\,\,\,\,\,\,\,\,\,\,\,\,\,
{\Box}^2\,\psi = 0\,.
\label{uravnilovki}
\eeq
The solutions of (\ref{uravnilovki}) can be presented
in the form
\beq
\ph = - 8\pi\sqrt{-b}\,\si + \ph_0\,,
\,\,\,\,\,\,\,\,\,\,\,\,\,\,\,\,\,\,\,\,\,
\psi =  \psi_0\,.
\label{reshen}
\eeq
where $\Box$ is the flat-space D'Alembertian 
and $\ph_0,\,\psi_0$ are general solutions of the
homogeneous equations
$\,{\Box}^2\,\ph_0=0\,,\,\,\,{\Box}^2\,\psi_0=0.\,$

There is an obvious arbitrariness related to the choice of the
initial conditions for the auxiliary fields $\ph,\,\psi$. 
However, replacing Eq. (\ref{reshen}) back into the action 
and taking variation with respect to $\si$ we arrive at the 
unique equation for $\sigma$. It proves useful to write this
equation in terms of $a(t)$ and the physical time $t$, with 
derivatives denoted by points.
Another useful variable is of course the Hubble parameter,
$H(t)= {\dot a}(t)/a(t) = {\dot \si}(t)$. 
Then, we obtain\footnote{We have included here $k=0,\pm 1$ for 
the sake of generality, but the rest of the paper will be only 
about the $k=0$ case.}
\beq
&& \frac{{\stackrel{....}{a}}}{a}
+\frac{{3\stackrel{.}{a}} {\stackrel{...}{a}}}{a^2}
+\frac{{\stackrel{..}{a}}^{2}}{a^{2}}
-\left( 5+\frac{4b}{c}\right) 
\frac{{\stackrel{..}{a}} {\stackrel{.}{a}}^{2}}{a^3}
-2k\left( 1+\frac{2b}{c}\right)
 \frac{{\stackrel{..}{a}}}{a^{3}}
\nonumber
\\
&& - \frac{M_{P}^{2}}{8\pi c}
\left( \frac{{\stackrel{..}{a}}}{a}+
\frac{{\stackrel{.}{a}}^{2}}{a^{2}}-\frac{2\Lambda }{3}\right)
\,=\,-\,\frac{\rho^0_m}{c\,a^3}\,.
\label{matter}
\eeq
In the {\it r.h.s.} of the last equation we have included the
contribution of matter with constant $\rho_m$ and explicitly 
shown dependence on $a(t)$. As far as we deal with the trace 
of the generalized Einstein equations (linear combination of 
generalized Friedmann equations), the contribution of radiation 
does not show up, but it can be easily restored if we switch to 
an equivalent $(00)$-component \cite{fhh,HHR,star}. 

There are several relevant observations we have to make about 
the solutions of Eq. (\ref{matter}) in different physical 
situations. 
First of all, in the theory without matter, when $\rho_m=0$, 
there are two exact solutions, namely 
\beq
a(t) \,=\, a_0 \cdot \exp(Ht)
\label{flat solution}
\eeq
where \cite{asta}
\beq
H\,=\, \frac{M_P}{\sqrt{-32\pi b}}\,\left(1\pm 
\sqrt{1-\frac{64\pi b}{3}\frac{\Lambda }{M_P^2}}\right)^{1/2}.
\label{H}
\eeq
As far as the cosmological constant is quite small compared 
to the square of the Planck mass, $\,\La \ll M_P^2$, we meet 
two very different values of $H$ (here $\La > 0$)
\beq
H_{c}\,\approx\,\sqrt{\frac{\Lambda }{3}}
\,\,\,\,\,\,\,\,\,\,\,\,
{\rm and}\,\,\,\,\,\,\,\,\,\,\,\,
H_S\,\approx\,\frac{M_P}{\sqrt{16\pi b}}\,.
\label{HH}
\eeq
It is easy to see that the first solution with $H_c$ is the 
one of the theory without quantum corrections, while the 
second value $\,H_S\,$ corresponds to the inflationary 
solution of Starobinsky \cite{star}. The sign of $\,b\,$ is 
always negative, independent on the particle content, see 
Eq. (\ref{b}). Let us remark that the particle content which
we deal with here, $\,N_0,\,N_{1/2}\,$ and $\,N_1\,$,
corresponds to the degrees of freedom contributing to the 
vacuum effective action and has nothing to do with the real 
matter content of the universe. 

Second, the stability properties of the solutions (\ref{HH}) 
depend on the sign of the coefficient $\,c$, that is on the 
coefficient of the local $\,\int\sqrt{-g}R^2$-term \cite{star,asta}. 
The inflationary solution $H_S$ is stable for a positive $\,c\,$ 
and is unstable for $\,c<0$. The stability of the low-energy 
solution $H_c$ requires opposite sign relations for $c$. It 
was shown in \cite{asta} that the $H_c$ solution with $\,c<0$ 
is stable with respect to the small variations of the Hubble 
parameter (or, equivalently, of $\si(t)$). In the original 
Starobinsky model of inflation \cite{star} the particles 
content corresponds to the unstable inflation and the initial 
data is chosen such that the Universe is asymptotically 
approaching the radiation-dominated FRW solution. 
In order to better understand the situation, let us replace 
the corresponding FRW solution, e.g., $a(t)\sim t^{2/3}$, 
into the equation (\ref{matter}). It is easy to see that the 
classical part, composed by Einstein and matter terms, do 
behave like $1/t^2$, while the quantum corrections, which 
are given by higher derivative terms, behave like $1/t^4$. 
This means that in the unstable phase the quantum terms do 
decay rapidly, such that the classical solution $a(t)\sim t^{2/3}$
is an excellent approximation to the solution of Eq. 
(\ref{matter}) in the corresponding epoch. It is an easy 
exercise to check that the same is true for the 
radiation-dominated $a(t)\sim t^{1/2}$ and cosmological
constant-dominated epochs too. An alternative form of these
considerations can be found in \cite{RadiAna}. 
Indeed, all the arguments presented above are valid 
for the dynamics of the conformal factor only. The next 
sections will be devoted to the stability of the same 
classical solutions with respect to the tensor metric 
perturbations (which we identify, for brevity, as gravitational 
waves). In the theory with quantum terms such as (\ref{vacuum}) 
and (\ref{finaction}), the equations for the gravitational 
wave have fourth derivatives and hence they represent a real 
danger for stability of the classical solutions. 

In the modified version of Starobinsky inflation 
\cite{Shocom,asta} the Universe starts with 
the stable inflation, that can be provided by choosing the 
supersymmetric particle content $\,N_0,\,N_{1/2}\,$ and $\,N_1\,$. 
Then the exponential inflation slows down due to the quantum 
effects of matter fields (mainly $s$-particles) and at some 
point the heavy $s$-particles decouple from gravity and then 
the Universe starts the unstable inflationary phase. The 
advantage of this version is that it does not depend on the 
choice of initial conditions. The stability of the stable 
inflation with respect to the metric perturbations will be 
explored in Sect. 6.

\section{Derivation of the gravitational waves} 

Let us derive the equation for the tensor modes of metric 
perturbations. First we rewrite the action in a more 
appropriate way and then derive linear perturbations for 
the tensor mode.

\subsection{Total action with quantum terms}

It proves useful to present the 
action (\ref{tota}) of a more useful way. After performing 
some integrations by parts, it can be cast into the form 
\beq
S\,=\,\int \,d^{4}x\,L\,,
\label{act}
\eeq
with
\beq
&& L \,=\,\sum^{5}_{s\,=\,0}\,f_{s}\,L_{s}
\label{1}
\\
\nonumber
&=& \sqrt{-g}\,\Big[f_{0}\,R 
+ f_{1}\,R^{\alpha \beta \mu \nu}R_{\alpha \beta \mu \nu} 
+ f_{2}\,R^{\alpha \beta}R_{\alpha \beta} 
+ f_{3}\,R^{2} 
+ f_{4}\,\ph\Box R 
+ f_{5}\,\varphi \Delta \varphi \Big]\,,
\eeq
where the $f$-terms are defined as
\beq
f_{0}\,&=&\,-\frac{{M_P}^{2}}{16\pi}\,;
\nonumber
\\
f_{1}\,&=&\,a_{1} + a_{2} - \frac{b + \omega}{2 
\sqrt{-b}}\,\varphi + \frac{\omega}{2 \sqrt{-b}}\,\psi;
\nonumber
\\
f_{2}\,&=&\,-2a _{1} - 4a_{2} 
+ \frac{\omega+2b}{\sqrt{-b}}\,\varphi 
- \frac{\omega}{ \sqrt{-b}}\,\psi;
\nonumber
\\
f_{3}\,&=&\,\frac{a _{1}}{3} + a_{2} - \frac{3c + 2b}{36} - 
\frac{3b + \omega}{6 \sqrt{-b}}\,\varphi + 
\frac{\omega}{6 \sqrt{-b}}\,\psi;
\nonumber
\\
f_{4}\,&=&\,-\frac{4\pi\sqrt{-b}}{3};
\nonumber
\\
f_{5}\,&=&\,\frac{1}{2}\,,
\label{fs}
\eeq
where the coefficients $a_{1,2}$ are the same defined in 
(\ref{vacuum}).

\subsection{Perturbation equations}

Using the conditions (with $\mu\,=\,0,1,2,3$ and $i\,=\,1,2,3$),
\beq
\partial_{i}\,h^{ij}\,=\,0\,\,\,\mbox{and}\,\,\,h_{kk}\,=\,0\,,
\eeq
together with the synchronous coordinate condition 
$h_{\mu 0}\,=\,0$, we introduce metric perturbation in the 
equation (\ref{1}) as follows
\beq
g_{\mu\nu}\,=\,g_{\mu\nu}^{0} + h_{\mu\nu}\,,\,\,\,\,\,
h_{\mu\nu}\,=\,\delta\,g_{\mu\nu}\,.
\eeq
Here $g_{\mu\nu}^{0}\,=\,\{1,-\delta_{ij}\,a^2(t)\}$ 
are the background cosmological solutions. In this way one can 
arrive at the following expressions for the bilinear parts of 
the partial Lagrangians from Eq. (\ref{1}):
\beq
L_{0}\,&=&\,a^{3}\,f_{0}\Big[h^{2}
\Big(\frac{3}{2}\dot{H} + 3 H^{2}\Big) 
+ h\ddot{h} + 4 H h \dot{h} + \frac{3}{4} \dot{h}^{2} - 
\frac{h}{4} \frac{\nabla^{2}h}{a^{2}}\Big] 
+ {\cal O}(h^{3}),
\nonumber
\\
L_{1}\,&=&\,a^{3}\,f_{1}\Big[{\dot{h}}^2\Big(2 H^{2} 
- 2\dot{H}\Big) - h\ddot{h}\Big(4 H^{2} 
+ 4\dot{H}\Big) - h^{2}\Big(3\dot{H}^{2} 
+ 6\dot{H} H^{2} + 6 H^{4}\Big) -
\nonumber
\\
\nonumber
&-& h\dot{h}\Big(8 H \dot{H} 
+ 16 H^{3}\Big) + \ddot{h}^2 + 4 H \dot{h}\ddot{h} + 
\Big(\frac{\nabla^{2}h}{a^{2}}\Big)^{2} 
+ 2\dot{h} \frac{\nabla^{2}\dot{h}}{a^{2}} + 
\\
&+& \Big(H^{2} h 
- 2 H \dot{h}\Big)\frac{\nabla^{2}h}{a^{2}}\Big] + {\cal O}(h^{3}),
\nonumber
\\
L_{2}\,&=&\,a^{3}\,f_{2}\Big[-h\dot{h}\Big(12\dot{H}H + 24 H^{3}\Big) 
- \frac{\dot{h}^{2}}{2}\Big(5\dot{H} + \frac{18}{4} H^{2}\Big) - 
\nonumber
\\
&-& h^{2}\Big(3\dot{H}^{2} + 9 \dot{H} H^{2} + 9 H^{4}\Big) -
h\ddot{h}\Big(4\dot{H} + 6 H^{2}\Big) + \frac{\ddot{h}^{2}}{4} 
+ \frac{3}{2} H \dot{h} \ddot{h} 
+  
\nonumber
\\
 &+&\frac{1}{4}\Big(\frac{\nabla^{2}h}{a^{2}}\Big)^{2} -
\frac{1}{2}\Big(\ddot{h} + 3 H \dot{h} - \dot{H} h - 
3H^{2} h\Big)\frac{\nabla^{2}h}{a^{2}}\Big] + {\cal O}(h^{3}),
\nonumber
\\
L_{3}\,&=&\,- 6 a^{3}\,f_{3}\Big(\dot{H} + 
2 H^{2}\Big)\Big[h^{2}\Big(\frac{3}{2}\dot{H} + 3 H^{2}\Big) 
+ 2 h \ddot{h} +
\nonumber
\\
&+& 8 H h \dot{h} + \frac{3}{2} \dot{h}^{2} 
- \frac{h}{2}\frac{\nabla^{2}h}{a^{2}}\Big] + {\cal O}(h^{3}),
\nonumber
\\
L_{4}\,&=&\,a^{3}\,f_{4}\,\Big[\frac{3}{2}\Big(\dot{H} 
+ 2 H^{2}\Big)\Big(\ddot{\varphi} + 3 H \dot{\varphi}\Big)h^{2} 
+ \Big(\ddot{\varphi} + 3 H \dot{\varphi}\Big) h \ddot{h} 
+ 
\Big(18 H^{2} \dot{\varphi} + 4 H \ddot{\varphi}\Big)h \dot{h} + 
\nonumber
\\
&+& 
\frac{3}{4}\Big(\ddot{\varphi} 
+ 3 H \dot{\varphi}\Big) \dot{h}^{2} 
- \frac{1}{4}\Big(\ddot{\varphi} 
+ 3 H \dot{\varphi}\Big)\frac{1}{a^{2}} h \nabla^{2} h\Big] 
+ {\cal O}(h^{3}),
\nonumber
\\
L_{5}
&=&
a^{3}\,f_{5}\,\Big\{\Big[\ddot{\varphi} h^{2} 
- \frac{3}{2} H \dot{\varphi} h^{2} - 
\dot{\varphi} \dot{h} h\Big] \ddot{\varphi} +
\nonumber
\\
&+& 
\dot{\varphi}^{2}\Big[-\frac{7}{4} H^{2} h^{2} 
- \frac{1}{4} \dot{H} h^{2} -\frac{7}{3} H \dot{h} h 
- \frac{1}{3} \ddot{h} h 
-\frac{1}{6} \frac{h}{a^{2}} \nabla^{2} h\Big]\Big\}
+ {\cal O}(h^{3}).
\label{Ls}
\eeq
One can perform a comparison of these equations with the ones
known from the literature. A very similar expansion was obtained 
by Gasperini in \cite{Gasperini:1997up} in order to explore metric 
perturbations in the pre-Big-Bang inflationary scenario. Despite 
the physical motivations of the pre-Big-Bang inflation are 
quite different from our case, when we intend to consider the 
semiclassical gravity with the action induced by anomaly, the 
formulas are mainly equivalent and we find a perfect correspondence 
between our results and the ones of \cite{Gasperini:1997up}. On 
the other hand, we can successfully compare part of the expressions 
presented above with our own previous calculation of the
same equations for the $H=const$ case in \cite{wave}. 

In order to get the equation for linearized tensor perturbations, 
one has to omit all higher order terms ${\cal O}(h^{3})$ in the 
expressions (\ref{Ls}) and then proceed by taking 
variational derivative with respect to $h_{\mu\nu}$. 
The next step is to use the solutions (\ref{reshen}) for the 
auxiliary fields. We fix the ambiguity in these solutions by 
choosing the simplest zero option for the conformal terms which 
are not controlled by conformal anomaly and set \ $\psi_0=\ph_0=0$.   
As a consequence we have 
\beq
\dot{\varphi}=-8\pi\sqrt{-b}\,H\,,\quad 
\ddot{\varphi}=-8\pi\sqrt{-b}\,{\dot H}\,,\quad 
\stackrel{...}{\varphi}=-8\pi\sqrt{-b}\,{\ddot H}
\quad \mbox{and}\quad
\stackrel{....}{\varphi}=-8\pi\sqrt{-b}\,\stackrel{...}{H}.
\label{ph and psi}
\eeq
These relations for the background must be replaced into the 
equations for tensor perturbations, but we prefer to keep 
$\ph$-dependent form, for the sake of generality. 
After all, the equation for tensor mode can be cast into the form
 \bea
&& \Big( 2 f_{1} + \frac{f_{2}}{2}\Big)\, \stackrel{....}{h}
+ \Big[ 3H\Big(4f_1 + f_2\Big) + 4 \dot{f_1} +
\dot{f_{2}}\Big]\, \stackrel{...}{h} 
+ \Bigl[ 3H^{2}\Big(6f_{1} + \frac{f_2}{2} - 4f_{3}\Big) 
\nonumber
\\
\nonumber
&+& H\Big(16 \dot{f_{1}} + \frac{9}{2}\dot{f_{2}}\Big)
+ 6\dot{H}\Big(f_1 - f_3\Big) + 2\ddot{f_{1}}
+ \frac{1}{2}\big(\ddot{f_2} + f_0 + f_4\ddot{\varphi}\big) 
+ \frac{3}{2} f_{4} H \dot{\varphi}
- \frac{2}{3} f_{5} \dot{\varphi}^{2}\Bigl]\,\ddot{h} 
\\
\nonumber
&-& \Big(4 f_{1} + f_{2}\Big)\,\frac{\nabla^2 \ddot{h}}{a^2}
+ \Big[ \dot{H}\Big(4\dot{f_{1}} - 6 \dot{f_{3}}\Big) 
- 21 H \dot{H} \Big(\frac{1}{2}f_2 + 2 f_3\Big) 
- \ddot{H}\Big(\frac{3}{2} f_2 + 6 f_3\Big) 
\\
\nonumber
&+& 3H^{2}\Big(4 \dot{f_{1}} + \frac{1}{2} \dot{f_{2}}
- 4 \dot{f_{3}}\Big) 
-9H^{3}\big(f_2 + 4f_3\big)
+ H \Big( 4\ddot{f_{1}} + \frac{3}{2}\ddot{f_{2}}\Big)
+ \frac{3}{2} f_4 \dot{\varphi}\Big(3 H^{2} + \dot{H}\Big) 
\\
\nonumber
&+& H \big( 3f_4 \ddot{\varphi} + \frac{3}{2} f_0
- 2 f_5\dot{\varphi}^2 \big)
+ \frac{1}{2} f_{4} \stackrel{...}{\varphi}
- \frac{4}{3} f_{5} \dot{\varphi} \ddot{\varphi} \Bigl]\, \dot{h} 
- \big[H\Big(4 f_{1} + f_{2}\Big) + 4 \dot{f_{1}}
+ \dot{f_{2}}\big]\,\frac{\nabla^{2} \dot{h}}{a^{2}}
\\
\nonumber
&+& \Big[5 f_{4} H \stackrel{...}{\varphi}
+ f_{4} \stackrel{....}{\varphi} 
- \big(36 \dot{H} H^{2} 
+ 18 \dot{H}^{2}   
+ 24 H \ddot{H}   
+ 4\stackrel{...}{H}\big) \Big(f_{1} + f_{2} + 3 f_{3}\Big) 
\nonumber
\\
\nonumber
&-& H \dot{H}\Big(32 \dot{f_{1}} + 36 \dot{f_{2}} +
120 \dot{f_{3}}\Big)
- 8\ddot{H}\Big(\dot{f_{1}} + \dot{f_{2}} + 3 \dot{f_{3}}\Big) 
- H^{2}\Big(4 \ddot{f_1} + 6 \ddot{f_2} + 24 \ddot{f_3}\Big)
\\
\nonumber
&-& 4\dot{H}\Big(\ddot{f_1} + \ddot{f_2} + 3 \ddot{f_3}\Big)
- 9f_{4}\dot{\varphi}\Big(H^3 + H \dot{H}\Big) 
+ f_{4} \ddot{\varphi}\Big(3 H^{2} + 5 \dot{H}\Big) 
- H^{3}\Big(8 \dot{f_{1}} + 12 \dot{f_{2}} + 48 \dot{f_{3}}\Big)
\\
\nonumber
&+& f_{5} \dot{\varphi}^{2}\Big(\frac{1}{2} H^{2}
+ \frac{1}{3} \dot{H}\Big) +
\frac{2}{3} f_{5} H \dot{\varphi} \ddot{\varphi}
- \frac{1}{6} f_{5} \ddot{\varphi}^{2} 
+ \frac{1}{3} f_{5} \dot{\varphi} \stackrel{...}{\varphi} \Bigl]\,h 
+ f_{0}\Bigl[2 \dot{H} + 3 H^{2}\Bigl]\,h
\\
\nonumber
&+& \Bigl[H^{2} \Big(4 f_{1} + 4 f_{2} + 12 f_{3}\Big)
+ H \Big(2 \dot{f_{1}} + \frac{1}{2} \dot{f_{2}}\Big)
+ 2\dot{H}\Big(f_1 + f_2 + 3 f_3\Big) 
\\
&-& \frac{1}{2}\, \big(\ddot{f}_2
+ f_4 \ddot{\varphi} + f_0  + 3 f_4 H \dot{\varphi} \big)
- \frac{1}{3} f_{5} \dot{\varphi}^{2}\Bigl]\,\frac{\nabla^{2} h}{a^{2}}
+ \Bigl[2 f_{1}
+ \frac{1}{2} f_{2}\Bigl]\,\frac{\nabla^{4} h}{a^{4}}
\,=\,0\,.
\label{diff}
\eea

Finally, if we take $H\,=\,constant$ in (\ref{diff}), we find 
the result which perfectly fits the one of \cite{wave}.

\section{Stability analysis}

Now we can start to deal with our main task and see whether 
Eq. (\ref{diff}) indicated that there is a stability of the 
cosmological solutions or not (similar research is done in 
\cite{Nelson:2010wp}). As a starting point we 
remember that a linear dynamical system with constant 
coefficients is stable when all poles, i.e., all roots 
of its characteristic equation have negative real part, 
i.e., they are on the left half complex plane. We can analyze 
this problem in two ways: numerically and analytically. 
We will start with an approximate analytical analysis.

\subsection{Semi-analytical analysis}

Let us start by rewriting the terms in (\ref{diff}) 
by using the plane wave representation in flat space section,
\beq
\frac{\nabla^{2} \ddot{h}}{a^{2}}
&=& - n^{2}\frac{\ddot{h}}{a(t)^{2}}
\,,\qquad
\frac{\nabla^{2} \dot{h}}{a^{2}}
\,=\,-n^{2}\frac{\dot{h}}{a(t)^{2}}
\,,
\nonumber
\\
\frac{\nabla^{2} {h}}{a^{2}}
&=&- n^{2}\frac{{h}}{a(t)^{2}}\,,\qquad
\frac{\nabla^{4} {h}}{a^{2}}\,=\,n^{4}\frac{{h}}{a(t)^{4}}\,.
\label{wave rep}
\eeq
Then the equation for tensor perturbations can be presented 
as follows
\beq
      b_{4}\,\stackrel{....}{h} 
\,+\, b_{3}\, \stackrel{...}{h} 
\,+\, b_{2}\, \stackrel{..}{h} 
\,+\, b_{1}\, \stackrel{.}{h} 
\,+\, b_{0}\, \stackrel{}{h} \,=\,0\,,
\label{mathematica}
\eeq
where we used the notations 
\beq
b_{4}\,&=&2 f_{1} + \frac{f_{2}}{2}\,
\label{b4}\,,
\\
b_{3}\,&=&\,3H\Big(4f_1 + f_2\Big) + 4 \dot{f_1} +
\dot{f_{2}}\,,
\label{b3}
\\
\nonumber
b_{2}\,&=&\,\Big(4 f_{1} + f_{2}\Big)\,\frac{n^{2}}{a^2} 
+ 3H^{2}\Big(6f_{1} + \frac{f_2}{2} - 4f_{3}\Big) 
+ H\Big(16 \dot{f_{1}} + \frac{9}{2}\dot{f_{2}}\Big)
+ 6\dot{H}\Big(f_1 - f_3\Big)
\\
&+&  2\ddot{f_{1}}
+ \frac{1}{2}\big(\ddot{f_2} + f_0 + f_4\ddot{\varphi}\big) 
+ \frac{3}{2} f_{4} H \dot{\varphi}
- \frac{2}{3} f_{5} \dot{\varphi}^{2}\,,
\label{b2}
\\
\nonumber
b_{1}\,&=&\,\big[H\Big(4 f_{1} + f_{2}\Big) + 4 \dot{f_{1}}
+ \dot{f_{2}}\big]\,\frac{n^2}{a^{2}} + \dot{H}\Big(4\dot{f_{1}} 
- 6 \dot{f_{3}}\Big) 
- 21 H \dot{H} \Big(\frac{1}{2}f_2 + 2 f_3\Big)  
\\
\nonumber
&-& \ddot{H}\Big(\frac{3}{2} f_2 + 6 f_3\Big) + 
3H^{2}\Big(4 \dot{f_{1}} + \frac{1}{2} \dot{f_{2}}
- 4 \dot{f_{3}}\Big) 
-9H^{3}\big(f_2 + 4f_3\big)
+ H \Big( 4\ddot{f_{1}} + \frac{3}{2}\ddot{f_{2}}\Big) 
\\
&+& \frac{3}{2} f_4 \dot{\varphi}\Big(3 H^{2} + \dot{H}\Big) + 
H \big( 3f_4 \ddot{\varphi} + \frac{3}{2} f_0
- 2 f_5\dot{\varphi}^2 \big)
+ \frac{1}{2} f_{4} \stackrel{...}{\varphi}
- \frac{4}{3} f_{5} \dot{\varphi} \ddot{\varphi}\,,
\label{b1}
\\
b_{0}\,&=&\, 5 f_{4} H \stackrel{...}{\varphi}
+ f_{4} \stackrel{....}{\varphi} 
- \big(36 \dot{H} H^{2} 
+ 18 \dot{H}^{2}   
+ 24 H \ddot{H}   
+ 4\stackrel{...}{H}\big) \Big(f_{1} + f_{2} + 3 f_{3}\Big) 
\nonumber
\\
\nonumber
&-& H \dot{H}\Big(32 \dot{f_{1}} + 36 \dot{f_{2}} +
120 \dot{f_{3}}\Big)
- 8\ddot{H}\Big(\dot{f_{1}} + \dot{f_{2}} + 3 \dot{f_{3}}\Big) 
- H^{2}\Big(4 \ddot{f_1} + 6 \ddot{f_2} + 24 \ddot{f_3}\Big)
\\
\nonumber
&-& 4\dot{H}\Big(\ddot{f_1} + \ddot{f_2} + 3 \ddot{f_3}\Big)
- 9f_{4}\dot{\varphi}\Big(H^3 + H \dot{H}\Big) 
+ f_{4} \ddot{\varphi}\Big(3 H^{2} + 5 \dot{H}\Big) 
\\
\nonumber
&-& H^{3}\Big(8 \dot{f_{1}} + 12 \dot{f_{2}} + 48 \dot{f_{3}}\Big) + 
f_{5} \dot{\varphi}^{2}\Big(\frac{1}{2} H^{2}
+ \frac{1}{3} \dot{H}\Big) +
\frac{2}{3} f_{5} H \dot{\varphi} \ddot{\varphi}
- \frac{1}{6} f_{5} \ddot{\varphi}^{2} 
+ \frac{1}{3} f_{5} \dot{\varphi} \stackrel{...}{\varphi}
\\
\nonumber
&+& f_{0}\Bigl[2 \dot{H} + 3 H^{2}\Bigl] - 
\Bigl[H^{2} \Big(4 f_{1} + 4 f_{2} + 12 f_{3}\Big)
+ H \Big(2 \dot{f_{1}} + \frac{1}{2} \dot{f_{2}}\Big)
+ 2\dot{H}\Big(f_1 + f_2 + 3 f_3\Big) 
\\
&+& \frac{1}{2}\, \big(\ddot{f}_2
+ f_4 \ddot{\varphi} + f_0  + 3 f_4 H \dot{\varphi} \big)
- \frac{1}{3} f_{5} \dot{\varphi}^{2}\Bigl]\,\frac{n^{2}}{a^{2}}
+ \Bigl[2 f_{1}
+ \frac{1}{2} f_{2}\Bigl]\,\frac{n^{4}}{a^{4}}\,.
\label{b0}
\eeq
The auxiliary field and its derivatives should be replaced 
according to Eq. (\ref{ph and psi}). 

So, we have to analyze the equation (\ref{mathematica}), 
with the coefficients \ $b_{k}$, \ ($k\,=\,0,...,3$) \ given 
by equations \ (\ref{b3}), \ (\ref{b2}), \ (\ref{b1}) \ and \
(\ref{b0}). One can easily reduce this fourth-order equation to 
a system of four first-order equations. 
Making the new change of variables we introduce 
\beq 
h_{0}\,=\,h\,,\quad 
h_{1}\,=\,\dot{h}_{0}\,=\,\dot{h}
\,,\quad 
h_{2}\,=\,\dot{h}_{2}\,=\,\ddot{h}
\,,\quad 
h_{3}\,=\,\dot{h}_{2}\,=\,\stackrel{...}{h}\,.
\label{h1234}
\eeq

Rewriting the differential equation, we arrive at
\bea
\nonumber
\dot{h}_{3}
&=& -\, \frac{1}{b_4}\,\Big(b_{3}\,{h}_{3} +
 b_{2}\,{h}_{2} + b_{1}\,{h}_{1} + b_{0}\,{h}_{0}\Big)\,,
\\
\nonumber
\dot{h}_{2}\,&=&\,h_{3}\,,
\\
\nonumber
\dot{h}_{1}\,&=&\,h_{2}\,,
\\
\nonumber
\dot{h}_{0}\,&=&\,h_{1}\,.
\eea

We can rewrite the linear system of four equations given above  
in a matrix form to compute its eigenvalues and eigenvectors. 
Thus, we can write in simplified form,
\beq
\dot{h}_{k}\,=\,A^l_k\,h_l\,,
\eeq
where $k\,=\,0,1,2,3$ and the matrix $A=A^l_k$ has the form
$$ 
A\,=\,\left(
\begin{array}{cccc}
0 & 1 & 0 & 0 \\
0 & 0 & 1 & 0 \\
0 & 0 & 0 & 1 \\
d_{0} & d_{1} & d_{2} & d_{3} \\
\end{array}
\right)\,, 
$$
Here we called \ $d_{k}\,=\,-b_{k}/b_{4}$\,.

If the coefficients of the matrix $A$ would be constants, the 
problem of stability could be solved immediately by deriving 
the eigenvalues of $A$. However, the same method is know to 
work even for the non-constant matrix $A$. The reason is that 
we are looking for the asymptotic stability related to the 
exponential time behavior of $h_k$. As far as the coefficients 
of the matrix $A$ have weaker time dependence, 
we can neglect this dependence and treat $A$ as a constant 
matrix. It is easy to see that this condition is 
satisfied for radiation and matter - dominated backgrounds and 
can be used also for the cosmological constant - dominated 
background, because $\La$ is actually very small. In other 
words, our constant-$A$ approximation means that we are 
looking for the dynamics of perturbations which is stronger 
than the expansion of the Universe in a given epoch. As far 
as we are interested in the consistency of classical cosmological 
solutions and in avoidance of dangerous run-away type 
solutions, this is definitely a very reliable 
approximation\footnote{The limitations of this analytical 
method are discussed, e.g., in \cite{Reed-Simon}.}. 

So, the next task is to find the eigenvalues of $A$ and hence
we consider
\beq
det\,\left(
\begin{array}{cccc}
-\lambda & 1 & 0 & 0 \\
0 & -\lambda & 1 & 0 \\
0 & 0 & -\lambda & 1 \\
d_{0} & d_{1} & d_{2} & (d_{3} -\lambda)\\
\end{array}
\right)\,=\,0\,.
\eeq
The algebraic equation
\beq
\lambda^{4} - d_{3}\,\lambda^{3} - d_{2}\,\lambda^{2} - 
d_{1}\,\lambda^{1} -d_{0}\,=\,0\,.
\label{polynomial}
\eeq
After some algebra (see Appendix for details), we can reduce 
the above equation to the following form 
\beq
z^{2} + \xi_{1}\,z + \xi_{2}\,=\,0\,.
\eeq
The most important quantity is 
\beq
\Delta\,=\,\xi_{1} + \,\frac{4}{27}\,\xi_{2}^{3}\,=\,
4\Biggl[\Bigl(\frac{\xi_{1}}{2}\Bigl)^{2} + 
\Bigl(\frac{\xi_{2}}{3}\Bigl)^{3}\Biggl]\label{delta}\,.
\eeq
The value of $ \Delta $, obtained by using the Cardano formula,
and all notations used here, are explained in Appendix.
Eq. (\ref{cardano}),  will tell us the nature of these roots. 
We can distinguish the following distinct cases: 
\begin{enumerate}
\item $\Delta\,<\,0$: Then the three roots are 
real and distinct and can be,
\begin{itemize}
\item All negative roots: stable.
\item Some positive root: unstable and instability generally
increases with increasing number of positive roots, in a 
sense one needs more severe initial conditions to avoid 
instability.
\end{itemize}
\item $\Delta\,=\,0$: the roots are real, 
and two or three are equal. Then,
\begin{itemize}
\item All negative roots or with negative real parts: stable.
\item Some root with a positive real part: unstable 
and this instability increase with increasing number 
of such positive roots.
\end{itemize}
\item $\Delta\,>\,0$: one real root and two complex roots,
\begin{itemize}
\item All negative roots or with negative real parts: stable.
\item Some root with a positive real part: unstable 
and this instability increase with increasing number 
of positive roots.
\end{itemize}
\end{enumerate}

In the case of equation (\ref{diff}) 
one meets the following values for (\ref{delta}),
\beq
\xi_{1} &=& \frac{-\alpha}{3} + \beta
\quad
\mbox{and}
\quad
\xi_{2}\,=\,\Biggl(\frac{2\alpha^{3}}{27} + 
\frac{3\gamma - \beta\,\gamma}{3}\Biggl)\,,
\nonumber
\\
\al &=& \frac{5}{2}\,p
\,;\quad
\gamma\,=\,\frac{1}{8}\Big(q^{2} - 4 p^{2} 
+ 4\,p\,r\Big)
\quad \mbox{and} \quad
\beta\,=\,2 p^{2} - r\,,
\label{condi}
\\
p &=& - \frac{39}{8}\,d_{3}^{2} + d_{2}\,;
\quad
q\,=\,\frac{d_{3}^{2}}{8} - \frac{d_{3}d_{2}}{2} + d_{1}
\quad
\mbox{and}
\quad
r\,=\,-\frac{3d_{3}^{4}}{256} + 
\frac{d_{2}d_{3}^{2}}{16} - \frac{d_{2}d_{1}}{4} + d_{0}\,.
\nonumber
\eeq
Let us remember that $b_{k}/b_{4}\,=\,-d_{k}$, where $b_{k}$ 
are given by (\ref{b3}), (\ref{b2}), (\ref{b1}), (\ref{b0}). 
Hence one can expect that the expression for $\Delta$ from 
the equation (\ref{delta}) will be rather complex, requiring 
numerical analysis. 

By performing such an analysis for the three cases of our 
interest, namely for exponential expansion, radiation and 
matter epochs, we find
\begin{enumerate}
\item 
Exponential expansion. 
When we choose $a_{1}\,<\,0$ we found $\Delta\,<\,0$. 
This is consistent because when we analyze equation 
(\ref{polynomial}) directly, we find all eigenvalues 
to be real and negative. So, we have the stability
in this case, exactly as we could expect from 
comparison to the inflationary case \cite{asta}.

However, if we choose \ $a_{1}\,>\,0$, then we find 
$\Delta\,<\,0$ too. But analyzing  equation (\ref{polynomial}) 
directly by numerical method (this means deriving the 
roots numerically by use of Mathematica software 
\cite{Wolfram}), we find three negative  and one 
positive eigenvalue. So, we can observe the instability 
in this case.

Let us remark that the sign of $a_1$ defines whether 
the massless tensor mode in the classical theory is 
a graviton or a ghost \cite{Stelle-1978}. From this 
perspective our result means that the stability 
property of the theory with higher derivative classical 
term (\ref{HD}) and quantum correction (\ref{nonloc})
is completely defined by classical part (\ref{HD}) and,
quite unexpectedly, does not depend on the quantum term
(\ref{nonloc}). Is it a general feature or just a 
peculiarity of the de Sitter background solution?
Let us consider other cases to figure this out. 

\item 
Radiation.
When we choose $a_{1}\,<\,0$ we found $\Delta\,>\,0$. 
This is consistent because when analyzing the Eq. 
(\ref{polynomial}) directly, we find two real eigenvalues, 
which are both negative and also two complex eigenvalues 
with negative real parts. So, we have stability in this 
case. But if we take $a_{1}\,>\,0$ it turns out that 
$\Delta\,<\,0$. Analyzing Eq. (\ref{polynomial}) 
directly, we find two negative eigenvalues 
and two positive ones. So, we have instability in 
this case. Again, the stability of the classical 
solution is completely dependent on the classical 
term (\ref{HD}). 

\item Matter. With $a_{1}\,<\,0$ we find $\Delta\,>\,0$. 
This is consistent the direct numerical analysis of Eq. 
(\ref{polynomial}), because in this way we find two real 
negative eigenvalues and also two complex eigenvalues with 
negative real parts. So, we have the stability for 
$a_{1}\,<\,0$. However, if we choose $a_{1}\,>\,0$, we 
find $\Delta\,>\,0$, indicating instability. 
By analyzing Eq. (\ref{polynomial}) 
directly one confirms this result, for we meet two 
real eigenvalues (one negative and other positive) and 
two complex ones, both with negative real parts. 
\end{enumerate}

As a result of our consideration we can conclude that 
there is a stability for Eq. (\ref{diff}), if and only 
if $a_{1}$ is negative. Taking into account the mentioned 
feature of classical higher derivative gravity, we see 
that the linear (in)stability of tensor mode in the 
classical higher derivative theory (\ref{HD}) completely 
defines a linear (in)stability in the theory with quantum 
correction (\ref{nonloc}). The qualitative explanation 
for this output is quite clear. The quantum terms 
(\ref{nonloc}) consist of two types of terms. The simplest 
one is the local $R^2$-term, which contributes to the 
propagator of gravitational perturbations on flat background, 
but not to the one of the tensor mode (see, e.g., \cite{book}
for detailed explanations and original references). The 
more complicated non-local terms are at least third order 
in curvature, and hence do not contribute at all to the 
propagator of gravitational perturbations on flat background.
Indeed, we are interested in the perturbations on curved 
cosmological background and not on the flat one. However, 
the typical length scale related to the expansion of the 
Universe is defined by the Hubble radius and are much 
greater than the length scale of the linear perturbations 
we are interested in here. Therefore, the stability of 
theory under such perturbations, in a given approximation, 
is the same as for the flat background and for the 
algebraic reasons explained above, there is no 
essential role of the anomaly-induced quantum terms  
(\ref{nonloc}) here.

\subsection{Numerical analysis}

In order to ensure that our qualitative and analytic 
consideration of the stability is correct, let us present 
the analysis of the stability of the differential equation 
(\ref{diff}) by means of numerical methods, using the software 
$Mathematica$ \cite{Wolfram}. 

We tested the same three relevant cosmological solutions, 
namely exponential expansion, radiation and matter. In all
cases the initial conditions of quantum origin were taken, 
for the sake of simplicity. For the mentioned three cases 
one can meet both stable and unstable solutions, as they
are shown in Figure 1 and Figure 2. 
One can easily see that the solutions where it was adopted 
$a_ {1} \, <\, 0$ are always stable, as shown in Figure 1. 
At the same time the solutions where it was adopted 
an opposite sign, $a_ {1} \,> \, 0$, are unstable, as shown 
in Figure 2.

\begin{figure}[H]
\centering
\includegraphics[scale=0.45]{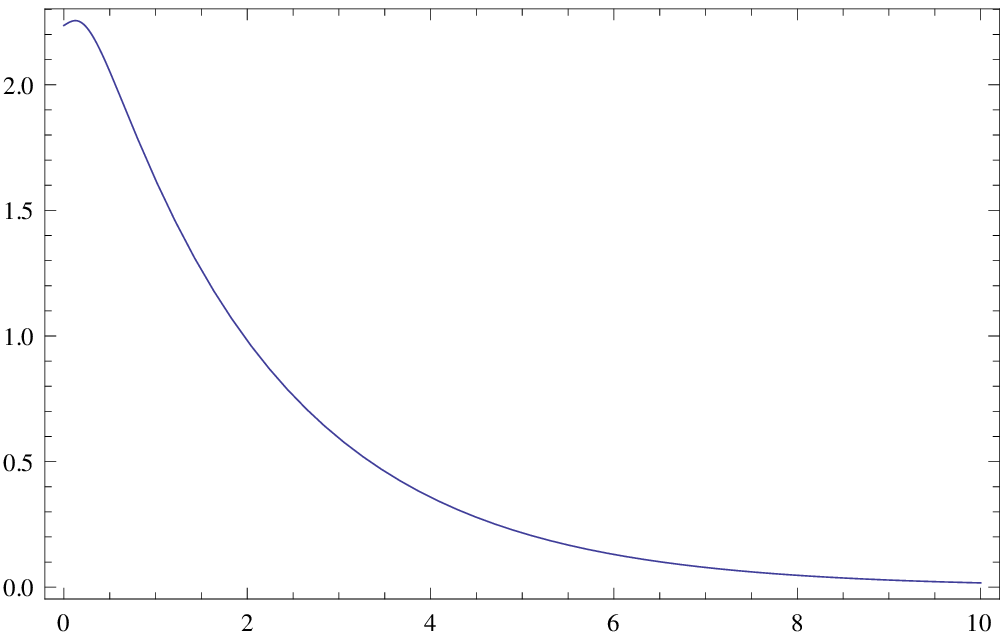}
\includegraphics[scale=0.45]{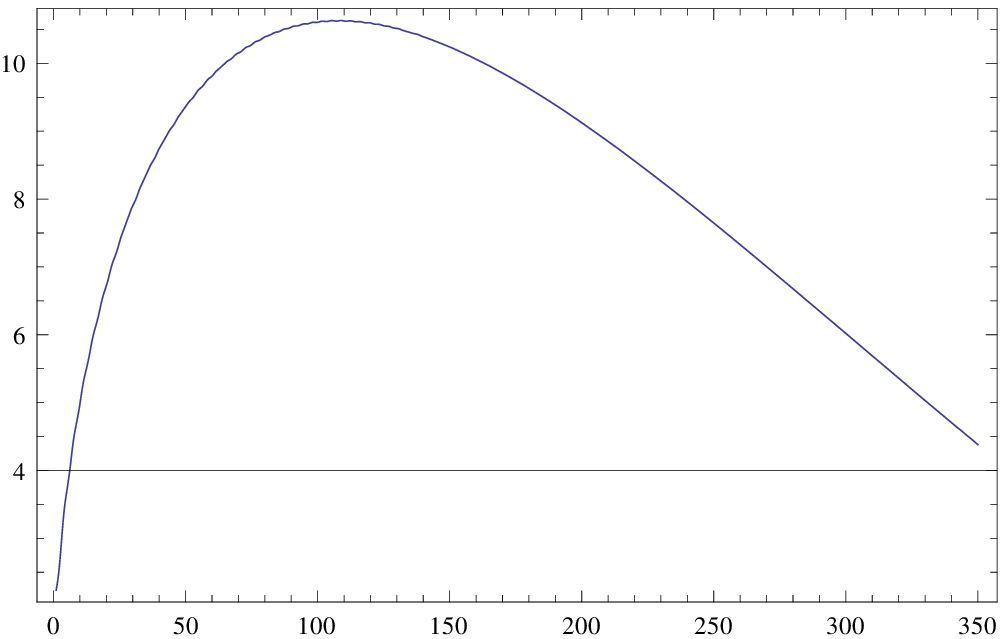}
\includegraphics[scale=0.45]{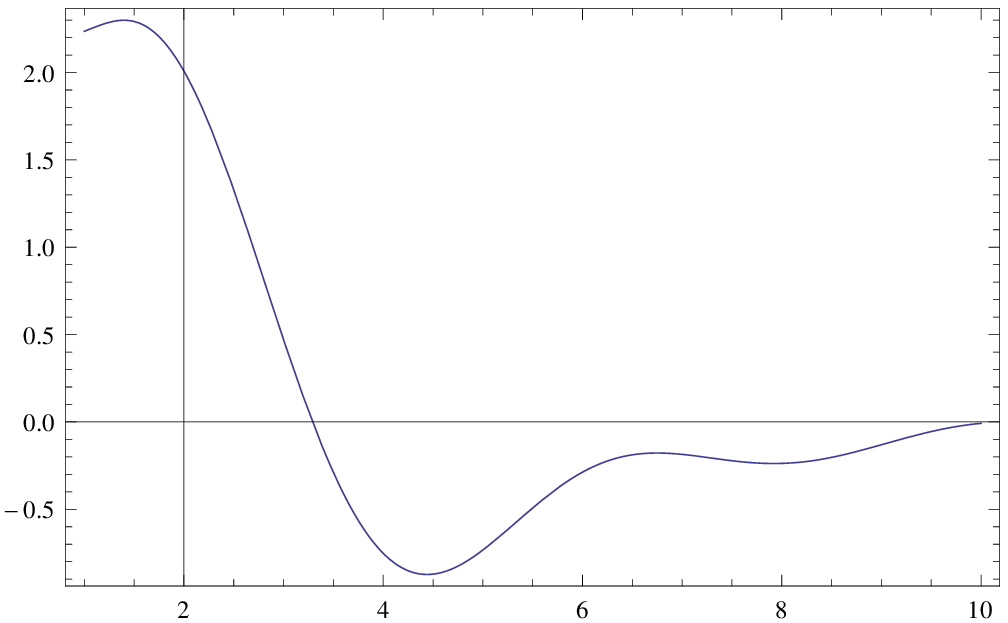}
\caption{Graph of perturbation $ h (t) $ in function of 
time analyzed in the cases for $a(t)\,=a_{0}e^{H_{0}t}$, 
$a(t)\,=a_{0} t^{1/2}$ and $a(t)\,=a_{0}t^{2/3}$ respectively, 
with the initial conditions $h_{0}\,=\,\frac{1}{\sqrt{2n}}$, 
$\dot{h}\,=\,\sqrt{\frac{n}{2}}$,$\ddot{h}
\,=\,\frac{n^{3/2}}{\sqrt{2}}$, 
$\stackrel{...}{h}\,=\,\frac{n^{5/2}}{\sqrt{2}}$, 
where we adopt $a_{1}\,<\,0$. Stable behavior.}
\label{figure1-1}
\end{figure}

\begin{figure}[H]
\centering
\includegraphics[scale=0.45]{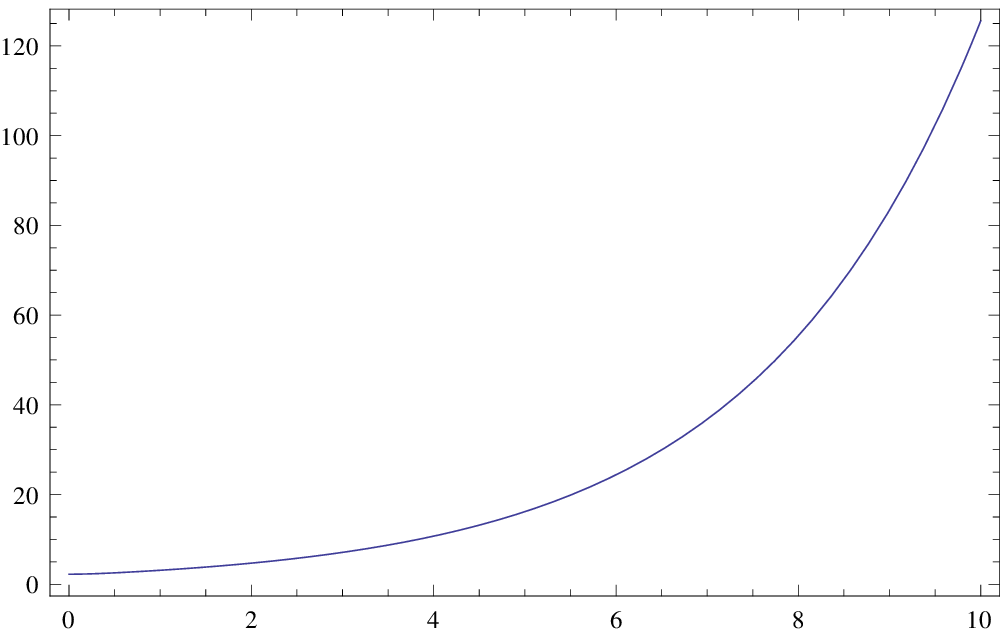}
\includegraphics[scale=0.45]{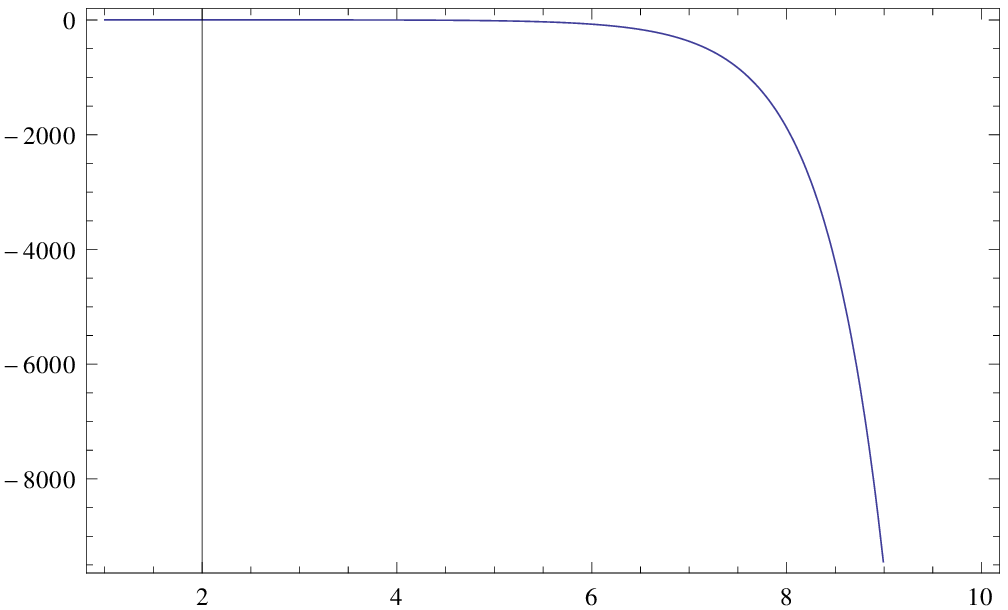}
\includegraphics[scale=0.45]{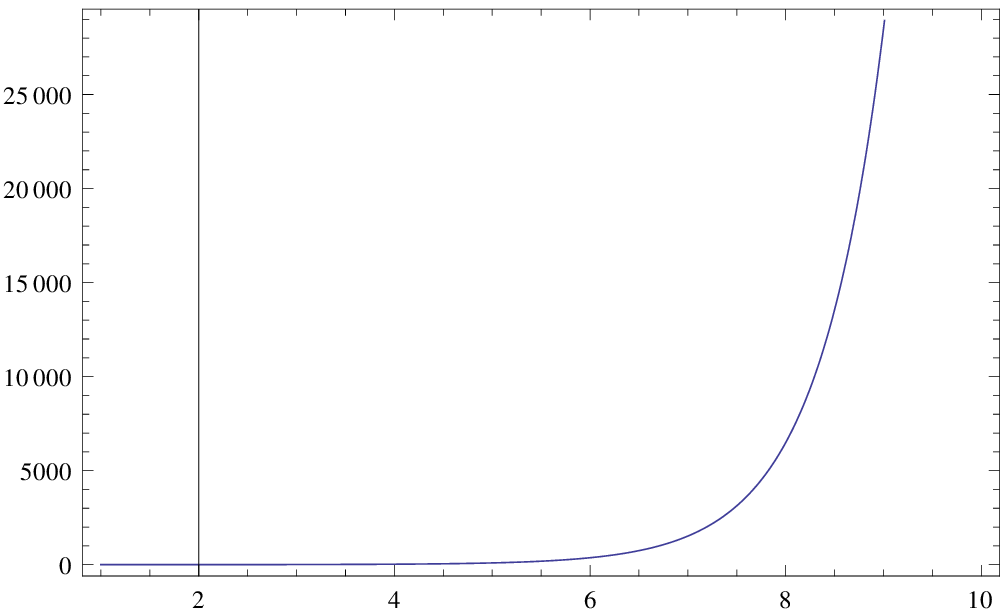}
\caption{Graph of perturbation $ h (t) $ in function of 
time analyzed in the cases for $a(t)\,=a_{0}e^{H_{0}t}$, 
$a(t)\,=a_{0}t^{1/2}$ and $a(t)\,=a_{0}t^{2/3}$ respectively, 
with the initial conditions $h_{0}\,=\,\frac{1}{\sqrt{2n}}$, 
$\dot{h}\,=\,\sqrt{\frac{n}{2}}$,$\ddot{h}
\,=\,\frac{n^{3/2}}{\sqrt{2}}$, 
$\stackrel{...}{h}\,=\,\frac{n^{5/2}}{\sqrt{2}}$, 
where we adopt $a_{1}\,>\,0$. Unstable behavior.}
\label{figure1-2}
\end{figure}

\section{Initial data problem and the spectrum of gravitational 
waves}

Here we develop a linearization process to found the spectral 
index $ k $ for our theory. But first we will test and fix our 
numerical method using the traditional inflation.

The perturbations, which originate from the fluctuations of the 
zero point energy of the quantum fields, have the spectrum 
characteristic of a scalar quantum field in Minkowski space. 
This ``vacuum state'' is well known \cite{birdav}:
\beq
h(x,\eta) = h(\eta)\,e^{\pm i\vec n.\vec x} \,,
\qquad 
h(\eta) \propto \frac{e^{\pm in\eta}}{\sqrt{2n}} \,.
\eeq
where we employed the conformal time, because the FRW metric
becomes conformal to the Minkowski metric in flat space; 
$\,\,\vec n$ is the wavenumber vector. A normalization 
constant was not considered in this expression. This constant 
is irrelevant for the determination of the spectral index, to 
be defined below but it is important to determine the full 
spectrum. After fixing this initial spectrum, one can derive 
how the initial amplitude depends on $\vec n$. In our case, 
it becomes
\beq
h_0 \propto \frac{1}{\sqrt{2n}} \,, \quad 
\dot h_0 \propto
\sqrt{\frac{n}{2}} \quad,
\quad \ddot h_0 \propto \frac{n^{3/2}}{\sqrt{2}} 
\,, \quad
{\stackrel{...} h}_0
\propto
\frac{n^{5/2}}{\sqrt{2}} \quad .
\label{initi}
\eeq
above the derivatives are taken with respect to the cosmic time, 
while the initial conditions are given in terms of the conformal 
time. The form of initial conditions tells us how they depend on 
the wavenumber value.

In order to study the dynamics of $h (t, \vec{x})$, it is 
necessary to make a Fourier transform,
\beq
h_{{\vec n}}(t)\,=\,\frac{1}{(2 \pi)^{3/2}}\,\int 
h(t,\vec{x})\,e^{ i {\vec n}\cdot \vec{x} }\,d^{3}x.
\eeq
We will need the total square of the amplitude, namely
\beq
h^{2}(t)\,=\,\int h_{{\vec n}}^{2}(t)\,d^{3}n\,.
\eeq
The above equation can be rewritten in the form
\beq
h^{2}(t)\,=\,4\pi \int h_{{\vec n}}^2(t)\,n^{2}\,d\,n
\,=\,
4\pi \int h^2_{{\vec n}}(t)\,n^3\,d\,\ln\,n
\,=\,4\pi\,\int P^{2}_{n}(t)\,d\,\ln\,n\,,
\eeq
where
\beq
P_{n}(t)\,=\,h_{{\vec n}}^{2}(t)\,n^3\,.
\eeq
The last quantity is called the ``power spectrum''. 
It shows us how the amplitude of gravitational waves vary in a 
range of $\,\ln\,n\,$ to $\,\ln (n + dn)$. 
Our definition follows closely the ones adopted in the 
book \cite{Weinberg-Cosmol}. Let us note that they are 
different from the ones used, e.g., in the Refs. 
\cite{mukha,Grishchuk:1993te}. It is important that the 
program CMBEASY \cite{CMBeasy} which we will use for numerical 
analysis, works with the same notations  \cite{Weinberg-Cosmol} 
which are adopted here.

Starting from this point we can find the power spectrum and the 
spectral index for the gravitational wave. For this end one has 
to square the value of the gravitational perturbation $h$ at a given 
time and for a given wave number $\,n$. We will vary this $\,n\,$ 
for a fixed $t$ and simultaneously solve our fourth-order 
differential equation (\ref{diff}) numerically. After this,
 we linearize the graph by plotting the relation
\beq
\ln n^3\,h_{{\vec n}}^{2}(t)\,\times\,\ln \,n\,.
\label{power}
\eeq
As a result we obtain the linear proportionality coefficient, 
which will be denoted as $\,k$ and called spectral index.  
Then we have, $\,P_{n}(t)\,\propto\,n^{k}$, i.e., it is 
proportional to the spectral index. It is the power spectrum 
that will tell us how the amplitude of the perturbations 
depends on the wavelength.

Now we will compare our results with the inflationary scenario based 
on the Einstein's equations with a cosmological constant. We can 
consider that this cosmological constant arise from some inflaton 
potential. So, one meets the deSitter
Universe, and the perturbations behave as \cite{Grishchuk:1993te},
\beq
h(\eta)  = \sqrt{\eta}\,g_\pm(n)H^{(1)}_{\pm\frac{3}{2}}(n\eta) \,,
\label{heta}
\eeq
where $g_\pm$ are integration constants which may depend on $n$, {and $H^{(1)}(x)$ is the Hankel function of first kind}.
This dependence must be fixed by the initial spectrum.
Taking the vacuum state as described above, we find that
those constants do not depend of $\,n$.
With the long wavelength limit approximation
$\,\,n \rightarrow 0\,\,$ and
considering the dominant mode in the above expression, we find
\begin{equation}
P_n \propto n^{3-3} \,=\, \mbox{constant} \,.
\end{equation}
The traditional inflationary scenario predicts a flat spectrum,
with $k = 0$. 

We can use these results to gauge our numerical procedure.
Fixing the initial spectrum according to (\ref{initi}),
integrating the equation for the
gravitational wave for the traditional inflationary scenario,
\begin{equation}
\ddot h\, -\, \frac{\dot a\,\dot h}{a}
\,+\, \biggr\{\frac{n^2}{a^2}
\,-\, 2\,\frac{\ddot a}{a}\biggl\}h \,=\, 0 \quad ,
\end{equation}
using $a(t) = e^{Ht}$, and the numerical
procedure described above, we found, for the (\ref{heta}) case,
\begin{equation}
k \simeq 0.01\label{regular-spectrum}\,.
\end{equation}
this numerical result is vary close to the analytical one, 
which is zero. We have considered a variation of $n$ between 
zero and one. Thus we find that to initial perturbations whose 
scales are of the order of the Planck length.

Our purpose is to apply the same procedure for our model, 
when the wave equation is given by Eq. (\ref{diff}). It is 
definitely not possible to find analytic solution of this 
equation, se one has to rely on numerical methods. The 
procedure implies that we find the dependence (\ref{power})
numerically and then extract the value of spectral index
$k$ in a way explained in the pedagogical example given 
above. In the case of exponential expansion without quantum 
corrections (this means $a_{1}\,=\,a_{2}\,=\,0$), we find
\beq
k\,=\,0.00894167\,\approx\,0.01\label{novo}\,.
\eeq
Let us additionally note that we have used the units with the 
Planck mass equal to one in the equation (\ref{diff}) and the 
Eqs. (\ref{fs}), in order to perform the numerical analysis.

Some remarks are in order. In the inflationary case without 
quantum corrections \cite{Hu:2000ti} the result is known to 
be $-0.15 < k < 0.16$. Therefore, the output of our numerical 
procedure is not really different from the ones of other 
models. This value (\ref{novo}) is very close to the one which 
can be found for inflation, where it is supposed to be 
$\,k \approx 0$, as we can see on (\ref{regular-spectrum}).
One can say that the model which we are dealing with  is 
similar to the well known inflationary case when we take 
$a_{1}\,=\,a_{2}\,=\,0$. Indeed, the flat or almost flat 
spectrum, which we have found, occurs because the stable version 
of the anomaly-induced inflation develops de Sitter phase with 
$\,H = const$.

Now we are able to analyze the general quantum case with
$a_{1}\,\neq\,0$. Let us remember that the value of $a_{1}$
can not be zero for it would violate the renormalizability. 
Moreover, the minimal possible magnitude of this parameter 
is defined by the magnitude of the corresponding $\be$-function
(\ref{w}) and hence can not be much less than $0.1$. Remember 
that the action for the vacuum is given by (\ref{vacuum}), 
where $\,a_1$ is most important for the evolution of tensor 
perturbations. Below we consider three separate cases and 
present the corresponding results of numerical analysis. 

Taking into account all the considerations made in sections 
2 and 4, above and also for the sake of generality, it is 
worthwhile to explore not only the case of inflation, but 
also radiation and matter - dominated epochs. In this way
we can compare the results of numerical procedure based on 
the CMBEASY program with the approximate analytic 
consideration presented in the previous section. 

\subsection{Inflation, or exponential expansion}

The variation of the power spectrum with respect to the 
change of $a_{1}$ (positive and negative) is presented in 
Table 1. 

\begin{table}[H]
\centering
\begin{tabular}{|c||c|c|c|c|c|c|c|c|c|c|c|}
\hline $a_{1}$ & 
0.0 & 0.1 & 0.2 & 0.3 & 0.4 & 0.5 & 0.6 & 0.7 & 0.8 & 0.9 & 1.0 
\\ 
\hline  $k$    & 
0.01 & 0.04 & 0.8 & 0.12 & 0.17 & 0.22 & 0.29 & 0.37 & 0.48 & 0.61 & 0.79 
\\ 
\hline 
\end{tabular} 
\caption{Getting $ k $ through the variation of $ a_{1}\,>\,0 $ 
in case of $a(t)\,=\,a_{0}\,e^{H_{0}t}$.}
\end{table}

\begin{table}[H]
\centering
\begin{tabular}{|c||c|c|c|c|c|c|c|c|c|c|c|}
\hline $a_{1}$ 
& 0.0 & -0.1 & -0.2 & -0.3 & -0.4 & -0.5 & -0.6 & -0.7 & -0.8 & -0.9 & -1.0 
\\ 
\hline  
$k$    
& 0.01 & -0.05 & -0.4 & -0.06 & -0.06 & -0.08 & -0.09 & -0.11 & -0.12 & -0.13 
& -0.14 
\\ 
\hline 
\end{tabular} 
\caption{Getting $ k $ through the variation of $ a_{1}\,<\,0 $ 
in case of $a(t)\,=\,a_{0}\,e^{H_{0}t}$.}
\end{table}

These two tables show that when the value of $\,a_1\,$ is negative 
and is decreasing, the values of $\,k\,$ also decrease while being 
negative. That is, as we increase $\,a_1$, the amplitude of 
gravitational waves decrease. In the case of positive $\,a_1\,$ 
an opposite situation happens, namely when we increase its value, 
the value of $\,k$ is also increasing. This happens because 
$\,a_1 \,> \,0\,$ does not represent a stable solution of 
(\ref{diff}).

These results may be compared with the computation of the spectral 
index for gravitational waves in the context of the pre-big bang 
scenario, based on the string effective action at tree level 
\cite{Gasperini-2003}. In this case, the spectral index is positive 
with the increasing spectrum, while the de Sitter inflation predicts 
a flat spectrum and power law inflation a negative spectral index, 
that means the decreasing spectrum.

Now, using the data obtained in \cite{Larson:2010gs}, 
\cite{Komatsu:2010fb}, \cite{Dunkley:2008ie} and using 
the software CMBEASY \cite{CMBeasy} 
\cite{Doran:2003sy} and \cite{Doran:2003ua} we can obtain the 
graph of the spectrum of anisotropy of the cosmic microwave 
background (CMB) due to gravitational waves for the cases 
specified above. In these graphs there are two essential 
quantities, namely the spectral index $\,k\,$ and the content 
of matter in the universe. The plots for the exponential 
expansion are shown in Figure 3. 
In all graphs we use $a_2\,=\,0$. However, 
$k$ value is the same for any $a_2$ value.

\FloatBarrier
\begin{figure}[H]
\center
\includegraphics[scale=0.4,angle=-90]{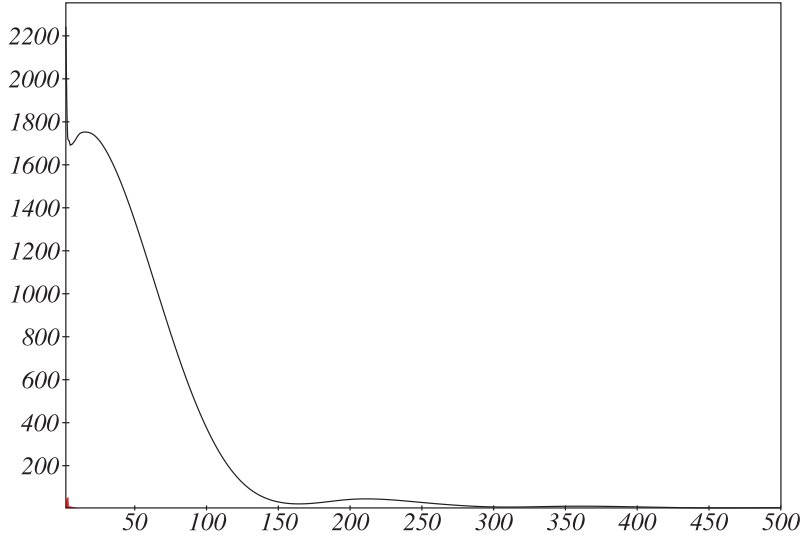}
\caption{Behavior of the anisotropy spectrum for the three cases with $k\,=\,0.01$. 
These three solutions have the same $\,k\,$ to $\,a_{1}=0$. That is, they are degenerate.}
\label{figure2}
\end{figure}

\subsection{Radiation}

By performing the same procedure of solving numerically the 
differential equation (\ref{diff}) and the same linearization 
as used before, we can test other cases, like radiation where 
we have $a(t)\,=\,a_{0}\,t^{1/2}$.

For radiation, we have the results presented in Table 3.

\begin{table}[H]
\centering
\begin{tabular}{|c||c|c|c|c|c|c|c|c|c|c|c|}
\hline $a_{1}$ & 
0.0 & 0.1 & 0.2 & 0.3 & 0.4 & 0.5 & 0.6 & 0.7 & 0.8 & 0.9 & 1.0 \\ 
\hline  $k$    & 
0.01 & 0.05 & 0.08 & 0.11 & 0.13 & 0.16 & 0.18 & 0.21 & 0.23 & 0.26 & 0.29 
\\ 
\hline 
\end{tabular} 
\caption{Getting $ k $ through the variation of $ a_{1}\,>\,0 $ 
in case of $a(t)\,=\,a_{0}\,t^{1/2}$.}
\end{table}

\begin{table}[H]
\centering
\begin{tabular}{|c||c|c|c|c|c|c|c|c|c|c|c|}
\hline $a_{1}$ & 
0.0 & -0.1 & -0.2 & -0.3 & -0.4 & -0.5 & -0.6 & -0.7 & -0.8 & -0.9 & -1.0 
\\ 
\hline  $k$    & 
0.01 & -0.26 & -0.27 & -0.28 & -0.33 & -0.30 & -0.33 & -0.38 & -0.38 & -0.36 
& -0.34 \\ 
\hline 
\end{tabular} 
\caption{Getting $ k $ through the variation of $ a_{1}\,<\,0 $ 
in case of $a(t)\,=\,a_{0}\,t^{1/2}$.}
\end{table}

\noindent
With these values we obtain the graphs shown in Figures 3 and 4, 
for the anisotropy spectrum, where we take the value \ $a_1=0$ 
\ (Figure 3) and the extreme value \ $a_1=1$ \ (Figure 4) from 
the tables presented above. One can see that the results of 
our numerical procedure are in accordance with the ones obtained
in Sect. 4 by the use of approximate analytical methods. In both 
cases the stability with the respect to the tensor modes holds 
for $a_1<0$ and only in this case.

\subsection{Matter}

Finally, we analyze the case for matter-dominated epoch, with 
$\,a(t)=a_{0}\,t^{2/3}\,$, where we get the values presented in 
Tables 5 and 6. 

\begin{table}[H]
\centering
\begin{tabular}{|c||c|c|c|c|c|c|c|c|c|c|c|}
\hline $a_{1}$ & 
0.0 & 0.1 & 0.2 & 0.3 & 0.4 & 0.5 & 0.6 & 0.7 & 0.8 & 0.9 & 1.0 \\ 
\hline  $k$    & 
0.01 & 0.06 & 0.09 & 0.12 & 0.14 & 0.17 & 0.20 & 0.22 & 0.25 & 0.28 & 0.29 
\\ 
\hline 
\end{tabular} 
\caption{Getting $\,k\,$ through the variation of $\,a_{1}>0\,$ 
in case of $\,a(t)=a_{0}\,t^{2/3}$.}
\end{table}

\begin{table}[H]
\centering
\begin{tabular}{|c||c|c|c|c|c|c|c|c|c|c|c|}
\hline $a_{1}$ & 
0.0 & -0.1 & -0.2 & -0.3 & -0.4 & -0.5 & -0.6 & -0.7 & -0.8 & -0.9 & -1.0 
\\ 
\hline  $k$    & 
0.01 & -0.23 & -0.24 & -0.24 & -0.28 & -0.25 & -0.28 & -0.32 & -0.31 & -0.29 
& -0.28 \\ 
\hline 
\end{tabular} 
\caption{Getting $\,k\,$ through the variation of $\,a_{1}<0\,$ 
in case of $\,a(t)=a_{0}\,t^{2/3}$.}
\end{table}

With these values we obtain the graphs for the anisotropy spectrum  
presented in Figures 3 ($a_1=0$) and 4 ($a_1=1$).  We can observe 
that the graphs of the spectrum of CMB anisotropy are nearly 
identical in the case of radiation and matter (Figure 3), but both 
are different from the case of inflation. The cases of matter 
and radiation in Figure 4 are degenerate.

\begin{figure}[H]
\center
\includegraphics[scale=0.5,angle=-90]{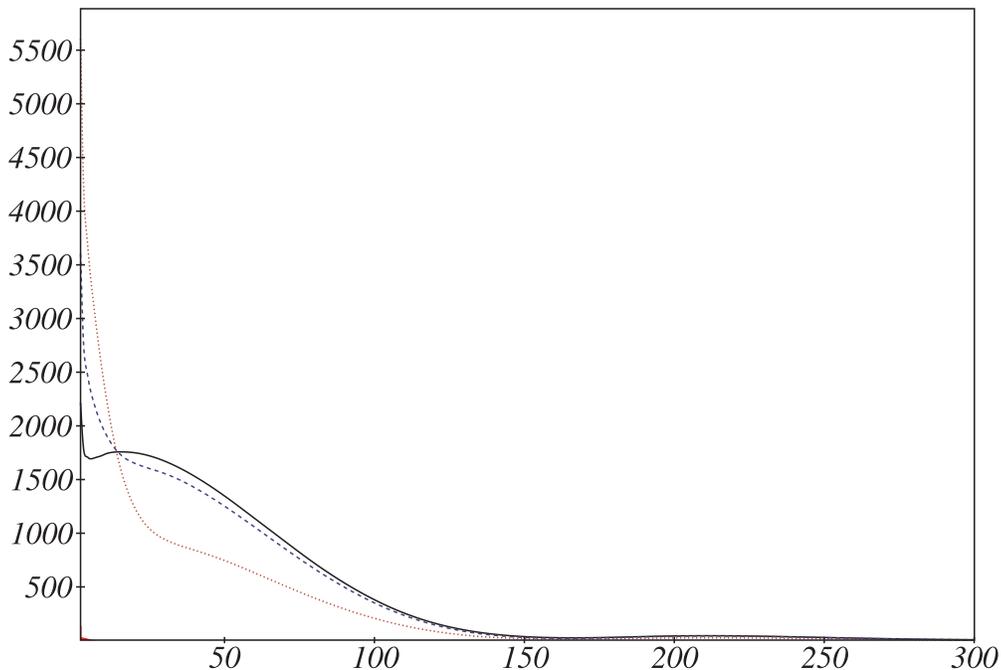}
\caption{Behavior of the anisotropy spectrum for gravitational 
waves in de Sitter (black line with $k\,=\,0$), inflation (red line 
with $k\,=\,0.79$) and Radiation/Matter (blue line with $k\,=\,0.3$ 
for both). All this is for $a_{1}\,=\,1$.}
\label{figure7}
\end{figure}

\noindent
Once again, we confirm the results obtained by approximate 
analytical methods in Sect. 4. Qualitatively the situation is 
pretty much the same as for the radiation case.

\section{Quantum effects of massive fields and tempered inflation}

It was shown in \cite{Shocom,asta} that taking the quantum 
effects of massive fields into account leads to the tempered 
form of inflationary solution $\,H_S\,$ in (\ref{HH}). The 
solution which emerges after we derive the corresponding 
effective equations can be very well approximated by the 
formula 
\beq
a(t) = e^{\sigma(t)}\,,
\qquad
\sigma(t)\,=\,H_{0}t - \frac{H_{0}^{2}}{4}\, \tilde{f}\,t^{2}\,,
\qquad
H(t)\,=\,H_{0} - \frac{H_{0}^{2}}{2}\,\tilde{f}\,t\,,
\label{massive}
\eeq
where $\tilde{f}$ is a small dimensionless parameter, which is 
at least as small as $10^{-5}$. The plot of this parabolic 
$\sigma(t)$ this given in Figure 5. This stable phase of inflation 
is supposed to last until $\,H(t)\,$ becomes comparable to the 
energy scale of supersymmetry breaking and the transition to 
the unstable phase of inflation occurs. 

\begin{figure}[H]
\center
\includegraphics[scale=0.9]{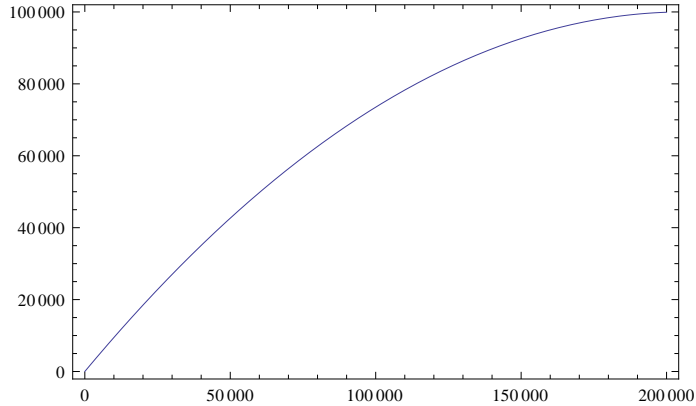}
\caption{Evolution of $\sigma(t)$.}
\label{figure8}
\end{figure}

We can use the solution (\ref{massive}) in our differential 
equation (\ref{diff}) for the tensor perturbation. Solving 
this equation numerically and doing linearization as it was 
explained above, we arrive at the Tables 7 and 8 for the 
variation of $\,k\,$ with respect to $\,a_1\,$ in the case 
of Modified Starobinsky model of inflation. 

\begin{table}[H]
\centering
\begin{tabular}{|c||c|c|c|c|c|c|c|c|c|c|c|}
\hline $a_{1}$ & 
0.0 & 0.1 & 0.2 & 0.3 & 0.4 & 0.5 & 0.6 & 0.7 & 0.8 & 0.9 & 1.0 
\\ 
\hline  $k$    & 
0.01 & 0.04 & 0.08 & 0.12 & 0.17 & 0.22 & 0.29 & 0.37 & 0.48 & 0.61 & 0.79 
\\ 
\hline 
\end{tabular} 
\caption{Getting $ k $ through the variation of $ a_{1}\,>\,0 $ in 
case of Eq. (\ref{massive}).}
\end{table}

\begin{table}[H]
\centering
\begin{tabular}{|c||c|c|c|c|c|c|c|c|c|c|c|}
\hline $a_{1}$ & 
0.0 & -0.1 & -0.2 & -0.3 & -0.4 & -0.5 & -0.6 & -0.7 & -0.8 & -0.9 & -1.0 
\\ 
\hline  $k$    & 
-0.01 & -0.02 & -0.03 & -0.04 & -0.05 & -0.07 & -0.09 & -0.11 & -0.12 & -0.13 
& -0.14 \\ 
\hline 
\end{tabular} 
\caption{Getting $k$ through the variation of $ a_{1}\,<\,0 $ in 
case of Eq. (\ref{massive}).}
\end{table}

We can see that the table is almost identical to the one for 
the case of usual inflation, described in Table 1. This is 
because the parameter $\,\tilde{f}\,$ is small ($\,10^{-5}\,$
in the numerical estimation which we used and many orders smaller 
in the realistic models, e.g., based on MSSM  \cite{Shocom,asta}), 
such that the typical time rate of slowing down inflation is much 
greater compared to the typical time scale of tensor perturbations. 
In other words, in the realistic cases the tensor modes behave 
very close to the case of purely exponential inflation. However, 
if we consider this parameter being increased, the two results 
begin to diverge from each other. The values of $\,k\,$ increase 
as the value of $\,\tilde{f}\,$ increase too.

\section{Conclusions}

The classical action of gravity which is necessary to guarantee 
the renormalizable theory of matter fields on classical metric
background goes beyond the conventional Einstein-Hilbert term, 
for it includes also the cosmological constant term and higher 
derivative terms. The presence of higher derivatives is usually 
associated to the problem of higher derivative unphysical ghosts, 
but this issue does not represent a real problem for the theory
when gravity is only an external background and there is no 
unitarity condition for the gravitational $S$-matrix. In the 
case of external gravity the necessary consistency conditions 
should be the existence of physically acceptable solutions and 
their robustness. The last criteria means that these solutions 
must be stable, at least, with respect to the small perturbations 
of the metric variables. In case of gravity the higher 
derivative terms can be especially dangerous for the stability 
of classical cosmological solutions, such as matter and 
radiation - dominated Universes and the Universe dominated 
by the cosmological constant term. The results found here 
can be seen as a first step in determining how the detection 
of the gravitational wave contribution to the spectrum of 
anisotropy of the cosmic microwave background radiation 
(which may be possible with the Planck satellite) may
lead to important bounds on the higher derivative terms 
and corresponding quantum corrections analyzed in this paper.

On the top of the classical vacuum terms there are also quantum 
corrections to it. One of the best available forms of such 
corrections is the anomaly-induced effective action of vacuum, 
which is especially efficient as an approximation to the 
present-day Universe, when all quantum fields except the photon 
have already decoupled from gravity and the quantum contribution
of photon fits perfectly to the anomaly-induced approach.  

We have used the anomaly-induced action of gravity to explore 
the behavior of tensor perturbations and especially to verify 
the stability of classical cosmological solutions with respect 
to such perturbations. The consideration was based on explicit 
derivation of gravitational wave equations in the theory with 
anomaly-induced quantum corrections and on the use of both 
analytical and numerical methods to perform the detailed analysis 
of these equations. The main conclusion of our work is that the 
stability conditions are essentially related to the sign of 
the Weyl-squared term in the {\it classical} action of vacuum 
and do not manifest any essential dependence on the quantum 
contributions. The qualitative explanation of this result is 
that the anomaly-induced action has a structure which prevents 
it to give essential contributions to the linear equation 
for tensor perturbations. Let us note that the situation may 
be very different for the density perturbations, where the sign 
of the $R^2$-term may be most relevant and for the non-linear 
perturbations, where the mentioned qualitative arguments 
simply do not hold. 

\section*{Acknowledgments.}

Authors are grateful to Grigory Chapiro 
for useful advises concerning the analytical method applied in 
Sect. 4.1. The work of J.F. has been partially supported by 
CNPq. The work of I.Sh. has been partially supported by CNPq, 
CAPES, FAPEMIG and ICTP. F.S. is thankful to FAPEMIG and CAPES 
for support. A.P. thanks CAPES/MEC-REUNI for financial 
support and DF/UFSC for hospitality.

\appendix 

\section{Roots of the quartic polynomial equation}

Here we present the analysis of the fourth-order polynomial 
which was used in Section 4. Consider the polynomial,
\beq
\lambda^{4} + d_{3}\lambda^{3} + d_{2}\lambda^{2} + 
d_{1}\lambda^{1} + d_{0}\,=\,0\,.
\eeq
Let us follow the classical Cardano method. 
To eliminate the cubic term make the substitution
$\lambda\,=\,x -\frac{d_{3}}{4}$. We get
\beq
x^{4} + p x^{2} + q x + r \,=\,0 \,,
\label{pol1}
\eeq
where
\begin{eqnarray}
p\,&=&\,\frac{d_3^2}{8} - \frac{d_3^2}{2} + d_2\,,
\\
q\,&=&\,\frac{d_3^2}{8} - \frac{d_3d_2}{2} + d_{1}\,,
\\
r\,&=&\,-\frac{3 d_3^4}{256} + \frac{d_2d_3^2}{16} - 
\frac{d_2d_1}{4} + d_0\,.
\end{eqnarray}
Now we rewrite (\ref{pol1}) as 
\begin{equation*}
x^{4} + p x^{2}\,=\, - q x - r 
\end{equation*}
by completing the square and adding the $y$ term. We obtain
\begin{equation}
y^{3} + \alpha y^{2} + \beta y + \gamma \,=\,0\,, 
\label{pol2}
\end{equation}
where
\begin{eqnarray}
\alpha\,&=&\,\frac{20}{8}\,p\,,
\\
\beta\,&=&\,2 p^{2} - r\,,
\\
\gamma\,&=&\,\frac{1}{8}(q^{2} - 4 p^{3} + 4 p r)\,.
\end{eqnarray}
In order to eliminate the second degree term in (\ref{pol2}), 
make one more substitution $y\,=\,z + m$, where $m\,=\,-\alpha/2$.
\beq
z^{3} + \xi_{1} z + \xi_{2}\,=\,0\,,
\label{pol3}
\eeq
where
\begin{eqnarray}
\xi_{1}\,&=&\,-\frac{\alpha}{3} + \beta\,,
\\
\xi_{2}\,&=&
\,\frac{2 \alpha^3}{27} + \frac{3 \gamma - \beta \gamma}{3}\,.
\end{eqnarray}
For the equation (\ref{pol3}) one can already apply 
the Cardano formula and find
\beq
\Delta\,=\,\xi_{1} + \,\frac{4}{27}\,\xi^{3}\,=\,
4\Biggl[\Bigl(\frac{\xi_{1}}{2}\Bigl)^{2} + 
\Bigl(\frac{\xi_{2}}{3}\Bigl)^{3}\Biggl]\,.
\eeq
The solution is
\beq
z\,=\,\sqrt[3]{-\frac{\xi_{1}}{2} 
+ \sqrt{\Bigl(\frac{\xi_{1}}{2}\Bigl)^{2} + 
\Bigl(\frac{\xi_{2}}{3}\Bigl)^{3}}} + \sqrt[3]{-\frac{\xi_{1}}{2} 
- \sqrt{\Bigl(\frac{\xi_{1}}{2}\Bigl)^{2} 
+ \Bigl(\frac{\xi_{2}}{3}\Bigl)^{3}}}\,,
\eeq
or
\beq
z\,=\,\sqrt[3]{-\frac{\xi_{1}}{2} + \frac{1}{2}\sqrt{\Delta}} 
+ \sqrt[3]{-\frac{\xi_{1}}{2} - \frac{1}{2}\sqrt{\Delta}}\,.
\label{cardano}
\eeq
As far as we know the values of $z$, $m$, $\alpha$ and $p$, 
it is possible to find the roots $\lambda$.
\vskip 6mm

\renewcommand{\baselinestretch}{0.9}


\end{document}